\documentclass[preprint,12pt]{elsarticle}

\usepackage{lineno,hyperref}
\usepackage{longtable}
\usepackage{pdflscape} 
\usepackage{subfig}
\usepackage{amsmath}
\usepackage{rotating}
\modulolinenumbers[200]

\journal{Journal of \LaTeX\ Templates}

\begin{document}

\begin{frontmatter}






\title{Systematic Trends of 0$^+_2$, 1$^-_1$,  3$^-_1$ and 2$^+_1$ Excited States in Even-Even Nuclei}
\cortext[cor1]{Corresponding author}
\author[label1]{B. Pritychenko\corref{cor1}}
\ead{pritychenko@bnl.gov}
\address[label1]{National Nuclear Data Center, Brookhaven National Laboratory, \\ Upton, NY 11973-5000, USA}

\author[label2]{B. Singh}
\address[label2]{Department of Physics $\&$ Astronomy, McMaster University, \\ Hamilton, Ontario, Canada L8S 4M1}

\author[label3]{M. Verpelli}
\address[label3]{Nuclear Data Section, International Atomic Energy Agency, \\ Vienna International Centre, P.O. Box 100, A-1400 Vienna, Austria}

\begin{abstract}
The spin and parity ($J^{\pi}$) assignments in even-even nuclei were reviewed across the nuclear chart using the Evaluated Nuclear Structure Data File (ENSDF). 
The prevalence of 2$^+_1$ first or lowest excited states is confirmed. The properties of   0$^+_2$, 1$^-_1$, and 3$^-_1$ lowest excited states were reexamined using the ENSDF data evaluation procedures. The $J^{\pi}$   systematic trends and correlations between level quantum numbers and nuclear physics phenomena are discussed.  
\end{abstract}


\end{frontmatter}

\linenumbers

\section{Introduction}


Comprehensive information on spin and parity assignments is essential for nuclear structure physics and model development. In the early 50s, Gertrude Scharff-Goldhaber at Brookhaven National Laboratory noticed the prevalence of 2$^+_1$ spin and parity assignments in even-even nuclei~\cite{Gold53}. 
Later, Beliaev and Zelevinsky at Kurchatov Institute explored this problem theoretically and proposed the nature of low-lying nuclear collective modes is just waves of the pair distortion~\cite{Bel62}. In well-deformed nuclei, these waves become  rotational states with angular momentum 2, but with a significantly lower moment of inertia than it would be for a macroscopic solid-body rotation as in a normal Fermi system. The proposed formalism described such fluctuations and quantized them as phonons plus anharmonicity that should be more pronounced than in macroscopic systems. Similar phenomena were investigated in macroscopic superconductors, the so-called Bardasis-Schriffer modes~\cite{Bar61,Sun20}.

It is well-established that the ground state spin and parity values  for even-even nuclei are 0$^+$~\cite{Mul00} while excited states assignments are less certain~\cite{ensdf,Dim20,NSDD}.  Historic compilations of the first excited states $J^{\pi}$ show the dominance of angular momentum 2 and positive parity~\cite{Gold53,Gold76,Boh69}, and these systematic trends have not been revisited since the 70s~\cite{Gold76,Gold78}. It is time to bridge the gap in the lowest excited state spin and parity trend assessments, update the compilations and study all presently-available data.

\section{ENSDF Survey of Even-Even Nuclei}

To evaluate spin and parity assignments across the nuclear chart, ENSDF relational database~\cite{ensdf,Dim20} was surveyed using Structured Query Language (SQL) queries at the Nuclear Data Section (NDS), International Atomic Energy Agency. The survey findings are shown in Table~\ref{table:dist}. 
\begin{table}[ht] 
\caption{ENSDF library~\cite{ensdf,Dim20} first excited states survey $J^{\pi}$ values.} 
\centering      
\begin{tabular}{l|c c}  
\hline\hline
$J^{\pi}$ & $\#$  & $\%$ \\
\hline
2$^+_1$   & 631 & 96.0 \\
0$^+_2$  & 20 & 3 \\
1$^-_1$ & 2 & 0.3 \\
3$^-_1$ & 2 & 0.3 \\
(8$^-_1$)?,(9$^-_1$,10$^-_1$)? & 2 & 0.3 \\
\hline\hline     
\end{tabular} 
\label{table:dist}  
\end{table} 


 The table data show that the first excited state spins and parities are known for 657 even-even nuclei, where 2$^+_1$ states are observed in 631 or 96$\%$ of even-even nuclei.  
Other observed $J^{\pi}$ values are 0$^+_2$, 1$^-_1$ and 3$^-_1$. Limited data for $^{254}$Rf and $^{270}$Ds suggest states with tentative spin and parity assignments of (8$^-$) and (9$^-$,10$^-$), respectively.  The level schemes in these two nuclei are fragmented and need further clarifications; they have to be reassessed in the future when more data will be available. 

 The 2$^+$ first-excited state assignment is natural for non-spherical rotational nuclei where the nuclear symmetry restricts the positive parity band to 0$^+$, 2$^+$, 4$^+$, ... states  
 \begin{equation}\label{eqn1}
J = 
\begin{cases}
  0, 2, 4, .... for & K^{\pi} = 0^{+} \\
  1, 3, 5, .... for &  K^{\pi} = 0^{-},
\end{cases} 
 \end{equation}
where $J$ is the total angular momentum, and $K$ is the projection of $J$ on the 3-axis in the intrinsic frame. In this case, 2$^+$  is the first excited positive parity band state, and its energy is defined by the minimum perturbation of the ground 0$^+$ state.  The energies of rotational states are described as
\begin{equation}\label{eqn2}
E_J = \frac{\hbar^2}{2I} J(J+1) + E_K,
\end{equation}
where $I$ is moment of inertia, and $E_K$ represents contributions from the intrinsic part of wave function. From formula~\ref{eqn2}, one can easily deduced a 10/3 ratio between 4$^+_1$ and 2$^+_1$ rotational states energies.  

 Spherical nuclei show a tendency for vibrational states:  2$^+_1$, 0$^+_2$, 2$^+_2$, 4$^+_2$, .... 
 Other lowest excited state assignments 0$^+$, 1$^-$ and 3$^-$ are found in closed-shell or subshell cases. 
Despite multiple efforts~\cite{Gold53,Bel62,Gold76}, we do not have a comprehensive nuclear theory that would describe first excited states in even-even nuclei  quantitatively~\cite{Zel21}, and nuclear data re-analysis can help to fill the void.

\section{Systematics of 0$^+_2$, 1$^{-}_1$, 3$^-_1$, and 2$^+_1$ Excited States}

In the present work, all available experimental data for non 2$^+$ low-lying states were critically reexamined using the standard ENSDF library procedures~\cite{Tul01}, and 2$^+$ data were adopted from the ENSDF library~\cite{ensdf,Dim20} and the dedicated horizontal evaluation of B(E2$\uparrow$)~\cite{Pri16}.  Numerical results for 0$^+_2$ and 1$^-_1$, 3$^-_1$ nuclear levels are shown in Tables~\ref{table:0+} and ~\ref{table:1-3-} while the complete list of 2$^+_1$ levels is given in Table~\ref{table:2+}.   Analysis of the  Tables~\ref{table:0+},~\ref{table:1-3-}, and ~\ref{table:2+} data implies that non 2$^+$ (0$^+$, 1$^{-}$ and 3$^-$)  lowest excited states occur near shell or subshell closure or in self-conjugate (N=Z)  nuclei.  Further examination of the  non 2$^+$   E4$^+_1$/E2$^+_1$ ratios produces numerical values from 1.058 to 2.072. These values  lie below the vibrational nuclei range of 2-2.2 and are completely inconsistent with the 10/3 ratio in rotational nuclei~\cite{Pri17}, and the nuclear shell model is needed for the interpretation of $J^{\pi}$  assignments in the previously mentioned nuclei.  Supplementary discussion on level properties is given in the following subsections.

\begin{landscape}
\begin{longtable}[c]{l c c c c c c c c} 
\caption{List of 0$^+_2$ Lowest Excited States in Even-Even Nuclei. Tentative spin and parity assignments are shown in parentheses. $\star$ - total width,   $\dagger$ - unconfirmed and *- single measurements. 
} \\
\hline\hline

\textbf{Nuclide} & \textbf{Z} & \textbf{N} & \textbf{J$^{\pi}$} & \textbf{Energy, keV} & \textbf{T$_{1/2}$}  & \textbf{Reaction/Decay}  & \textbf{Remarks}  & \textbf{E4$^+_1$/E2$^+_1$}  \\
\hline 
\endfirsthead
\hline
\textbf{Nuclide} & \textbf{Z} & \textbf{N} & \textbf{J$^{\pi}$} & \textbf{Energy, keV} & \textbf{T$_{1/2}$}  & \textbf{Reaction/Decay}  & \textbf{Remarks}  & \textbf{E4$^+_1$/E2$^+_1$}  \\
\hline 
\endhead
\hline
\multicolumn{9}{r}{\textit{Continued on next page ...}} \\
\endfoot
\endlastfoot

$^{4}$He & 2 & 2 & 0+ & 20100 (50) &  0.27(5) MeV$^{\star}$   & $^{4}$He(e,e') & \cite{1970Wa24} & \\
               &  2 & 2 & 0+ & $\sim$20000 &                 & $^{4}$He(e,e') & \cite{1968Fr04}   &  \\
$^{12}$O & 8 & 4 & 0+ & 1620 (110) & 1.2(7) MeV$^{\star}$  & $^1$H($^{14}$O, t) & \cite{2016Su05} & \\
$^{16}$O & 8 & 8 & 0+ & 6049.4 (10) &  & $^{16}$O(e,e') & \cite{1975Mi08}  & 1.497 \\
                & 8 & 8 & 0+ & 6049.4 (10) & 67(5) ps & $^{19}$F(p,$\alpha$)  & \cite{1973Bi17} &  \\
$^{44}$Ar$^{\dagger}$ & 18 & 26 & (0+)? & 750 (30) &  & $^{48}$Ca($^{3}$He,$^{7}$Be) &  J$^{\pi}$ from shell model~\cite{1976Cr03} & 2.371 \\
$^{40}$Ca & 20 & 20 & 0+ & $\sim$3350 & 2.15 (8) ns  & $^{40}$Ca(p,p')  & \cite{1966Go23} & 1.352 \\
                 & 20 & 20 & 0+ & $\sim$3350 & 2.14 (10) ns  & $^{40}$Ca(n,n') & \cite{1973Ba69} & \\
                & 20 & 20 & 0+ & 3352.62 (9) &           & $^{40}$Ca(p,p')  & L=0, \cite{1993Se02} & \\
$^{68}$Ni & 28 & 40 & 0+ & 1604.0(4) keV &   & $^{68}$Co($\beta^{-}$) & \cite{2015Fl01,2014Su05} & 1.548 \\
                & 28 & 40 & 0+ &  &  270 (5) ns  & $^{58}$Ni($^{70}$Zn,X$\gamma$)  & \cite{2002So03} &  \\
$^{72}$Ge & 32 & 40 & 0+ & 688 (3) &   & $^{70}$Ge(t,p) & L=0, \cite{1979Mo08}  & 2.072 \\
                  & 32 & 40 & 0+ &  &    &  $^{74}$Ge(p,t) & L=0, \cite{1982Be13,2007Fr10} &  \\
                  & 32 & 40 & 0+ &   &  444.2 (8) ns  & $^{72}$Ge(n, n'$\gamma$) & \cite{1984Br24} &  \\
$^{72}$Kr$^*$ & 36 & 36 & 0+ & 671.0 (10) &  26.3(21) ns  &  $^{9}$Be($^{78}$Kr,X)  & \cite{2003Bo05} & 1.860 \\
$^{90}$Zr & 40 & 50 & 0+ & 1761  &  & $^{92}$Zr(p,t) & L=0, \cite{1971Ba43} & 1.407 \\
                 & 40 & 50 & 0+ & 1760 & 61.3 (25) ns  & $^{90}$Zr(p,p'$\gamma$) & \cite{1972Bu18}  &  \\
                & 40 & 50 & 0+ & 1760.72 &  & $^{90}$Y($\beta^{-}$) & Branching~\cite{2020Pi01,2020Dr01} &  \\
$^{96}$Zr & 40 & 56 & 0+ & 1581.4 &   &  $^{96}$Y($\beta^{-}$) & Conv. data~\cite{1990Ma03} & 1.571 \\
                & 40 & 56 & 0+ & 1594 (8) &  & $^{94}$Zr(t,p)  & L=0, \cite{1974Fl02} &  \\
               & 40 & 56 & 0+ & 1590 &  38.0 (7) ns & $^{96}$Zr(p,p'$\gamma$)   & \cite{1972Bu18} &  \\
$^{98}$Zr & 40 & 58 & 0+ & 854 &  & $^{98}$Y($\beta^{-}$)  &  Conv. data~\cite{1994Lh01} & 1.507  \\
                & 40 & 58 & 0+ & 853 & 63 (7) ns  & $^{98}$Y($\beta^{-}$)  & \cite{1977Si05} &  \\
                & 40 & 58 & 0+ &   & 65 (10) ns  & $^{235}$U(n,F$\gamma$) & \cite{1976Po11} &  \\
$^{98}$Mo & 42 & 56 & 0+ & 734 &  &  $^{96}$Mo(t,p) & Conv. data~\cite{1984De17}  & 1.919  \\
                   & 42 & 56 & 0+ & 735 (5) & & $^{96}$Mo(t,p)  & L=0, \cite{1981Fl06} &  \\
                  & 42 & 56 & 0+ & 737 &  & $^{100}$Mo(p,t)  & L=0, \cite{1973Sh09} &   \\
                   & 42 & 56 & 0+ &  & 21.8 (9) ns & $^{98}$Mo(p,p'$\gamma$) & \cite{1972Bu18} &   \\
$^{180}$Hg & 80 & 100 & 0+ & 419.6 &   & $^{180}$Tl(EC)  & Conv. data \cite{2011El07} & 1.627 \\
                   & 80 & 100 & 0+ & 420 &   & $^{147}$Sm($^{36}$Ar,3n$\gamma$) & Conv. data~\cite{2011Pa24}  &  \\
$^{182}$Hg & 80 & 102 & 0+ & 335 (1) &  & $^{182}$Tl(EC)  & Conv. data~\cite{2017Ra11}  & 1.741 \\
$^{184}$Pb & 82 & 102 & (0+) & 572 (30) &  & $^{188}$Po($\alpha$) & Low hindr.~\cite{2003Va16} & \\
                   & 82 & 102 & (0+) & 577 (40) &  & $^{188}$Po($\alpha$) & \cite{1999An52} & \\
$^{186}$Pb & 82 & 104 & (0+) & 532 &  & $^{190}$Po($\alpha$) &  Low hindr.~\cite{2000An14} & 1.394 \\
$^{188}$Pb & 82 & 106 & 0+ &  &  &  $^{192}$Po($\alpha$)  & Low hindr. ~\cite{2003Va16} & 1.470 \\
                   & 82 & 106 & 0+ & 591 (2) &  & $^{156}$Gd($^{36}$Ar,4n$\gamma$) & Conv. data~\cite{1999Le61}  &  \\
$^{190}$Pb & 82 & 108 & 0+ & 658 (4) &  & $^{194}$Po($\alpha$)  &  Low hindr. ~\cite{1989De18,1994Wa13} & 1.588 \\
                    & 82 & 108 & 0+ & 658 & $\leq$0.22 ns & $^{194}$Po($\alpha$)  & Conv. data~\cite{1989De18}  &  \\
$^{192}$Pb & 82 & 110 & 0+ & 768.5 (4) &  & $^{192}$Bi(EC) & Conv. data~\cite{1987Va09,1990Tr01} & 1.587  \\
                   & 82 & 110 & 0+ & 768.5 (17) & 0.75 (10) ns & $^{196}$Po($\alpha$)  & \cite{1989De18} &  \\
                    & 82 & 110 & 0+ &  &  &  $^{196}$Po($\alpha$)  & Low hindr.~\cite{1985Va03}  &   \\
$^{194}$Pb & 82 & 112 & 0+ & 930.6 (4) &  & $^{194}$Bi(EC) & Conv. data~\cite{1987Va09,1990Tr01}  & 1.596 \\
                   & 82 & 112 & 0+ & 931 &  &  $^{198}$Po($\alpha$) & Low hindr.~\cite{1989De18,1994Wa13}  &  \\
                   & 82 & 112 & 0+ & 930.6 (9) & 1.1 (2) ns &  $^{198}$Po($\alpha$)  & Conv. data~\cite{1989De18} &  \\
\hline\hline

\label{table:0+}  
\end{longtable}
\end{landscape}


\begin{landscape}
\begin{longtable}[c]{l c c c c c c c c} 
\caption{List of 1$^-_1$, 3$^-_1$  and Others Lowest Excited States in Even-Even Nuclei. Tentative spin and parity assignments are shown in parentheses.  $\dagger$-$^{254}$Rf and $^{270}$Ds tentatively-assigned levels (8-) and (9-,10-), respectively. The listed levels may not be the first excited states as not much is known about the level structures of these nuclei.
} \\
\hline\hline

\textbf{Nuclide} & \textbf{Z} & \textbf{N} & \textbf{J$^{\pi}$} & \textbf{Energy, keV} & \textbf{T$_{1/2}$}  & \textbf{Reaction/Decay}  & \textbf{Remarks}  & \textbf{E4$^+_1$/E2$^+_1$}  \\
\hline 
\endfirsthead
\hline
\textbf{Nuclide} & \textbf{Z} & \textbf{N} & \textbf{J$^{\pi}$} & \textbf{Energy, keV} & \textbf{T$_{1/2}$}  & \textbf{Reaction/Decay}  & \textbf{Remarks}  & \textbf{E4$^+_1$/E2$^+_1$}  \\
\hline 
\endhead
\hline
\multicolumn{9}{r}{\textit{Continued on next page ...}} \\
\endfoot
\endlastfoot

$^{14}$C & 6 & 8 & 1- & 6093.8( 2) &  & $^{12}$B($\beta^{-}$) & \cite{2016Fi06}  & 1.531 \\
                & 6 & 8 & 1- &  &  &  $^{14}$C($\alpha$,$\alpha$') & L($\alpha$,$\alpha$')=1, \cite{1984Pe24} &  \\
                & 6 & 8 & 1- &  & $\leq$7 fs & $^{9}$Be($^{13}$C,$^{8}$Be) & \cite{1975Se04} &  \\
$^{14}$O & 8 & 6 & 1- & 5164 (2) &  & $^{9}$Be($^{15}$O,$^{14}$O) & \cite{2019Ch50} &  1.505  \\
                & 8 & 6 & 1- &  &   & $^{12}$C($^{3}$He,n) & L=1, \cite{1987Ab04} &    \\
                & 8 & 6 & 1- &  & 38.1 (18) keV& $^{14}$N($^{3}$He,t)  & \cite{1985Ch06} &   \\
$^{146}$Gd & 64 & 82 & 3- & 1579.40 (5) &   & $^{144}$Sm($\alpha$, 2n) & GTOL~\cite{Tul01} &  1.325 \\
                    & 64 & 82 & 3- &  &   & $^{144}$Sm($\alpha$, 2n) & J$^{\pi}$~\cite{2010Ca08,1986Ya06} &  \\
                    & 64 & 82 & 3- & &   & $^{146}$Tb(EC) & Conv. data~\cite{1995GoZV} &   \\
                    & 64 & 82 & 3- & &  1.06 (12) ns & $^{144}$Sm($\alpha$, 2n) & \cite{1978Kl04} &   \\
$^{208}$Pb & 82 & 126 & 3- & 2614.511 (10) &  &  & Evaluation~\cite{2000He14} &  1.058 \\
                   & 82 & 126 & 3- &  & 16.7 (3) ps & $^{208}$Pb($^{16}$O, $^{16}$O') & B(E3)~\cite{1983Sp02} &   \\
                   & 82 & 126 & 3- &  &  & $^{208}$Pb($\alpha$,$\alpha$') & L($\alpha$,$\alpha$')=3, \cite{1992Fu01} &  \\
                   & 82 & 126 & 3- &  &  & $^{208}$Pb(p,p') & L(p,p')=3, \cite{1980Ad01} &  \\
                   & 82 & 126 & 3- &  &  & $^{208}$Pb(e,e') & L(e,e')=3, \cite{1968Zi02}  &  \\
$^{254}$Rf$^{\dagger}$ & 104 & 150 & (8-) & $>$1350 keV & 4.7 (11) $\mu$s  &  & \cite{2015Da12}  &  \\
$^{270}$Ds$^{\dagger}$  & 110 & 160 & (9-,10-) & 1.13E+03 & 3.9$^{+15}_{-8}$ ms &  & \cite{2017Ac02, 2015Ac04} &  \\
\hline\hline     
\label{table:1-3-}  

\end{longtable}
\end{landscape}

\subsection{ 0$^+_2$  Excited States}

The low-lying 0$^+$ states in even-even nuclei serve as an indication of shape coexistence~\cite{Mor56,Hey11}.  
Analysis of the Table~\ref{table:0+} data demonstrates that 0$^+$ first excited states materialize near Z, N=2,8,20, N=40, and Z=82, often in self-conjugate nuclei. The two notable exceptions are $^{44}$Ar~\cite{1976Cr03,Che11} and $^{72}$Kr~\cite{2003Bo05,Abr10}. Further examination of $^{44}$Ar~\cite{Sch96} raises serious doubts about the existence of a low-lying 0$^+$ level at 750 keV due to background spectra interpretations. The situation with the $^{72}$Kr 0$^+$ state is more complex because spin and parity assignments are based on a single measurement~\cite{2003Bo05,Abr10}. Likely, the 0$^+_2$   state in  $^{44}$Ar  was erroneously reported in ENSDF while $^{72}$Kr needs additional measurements.

\begin{figure}
  \subfloat[]{%
  \begin{minipage}{\linewidth}
\centering
  \includegraphics[width=.6\linewidth]{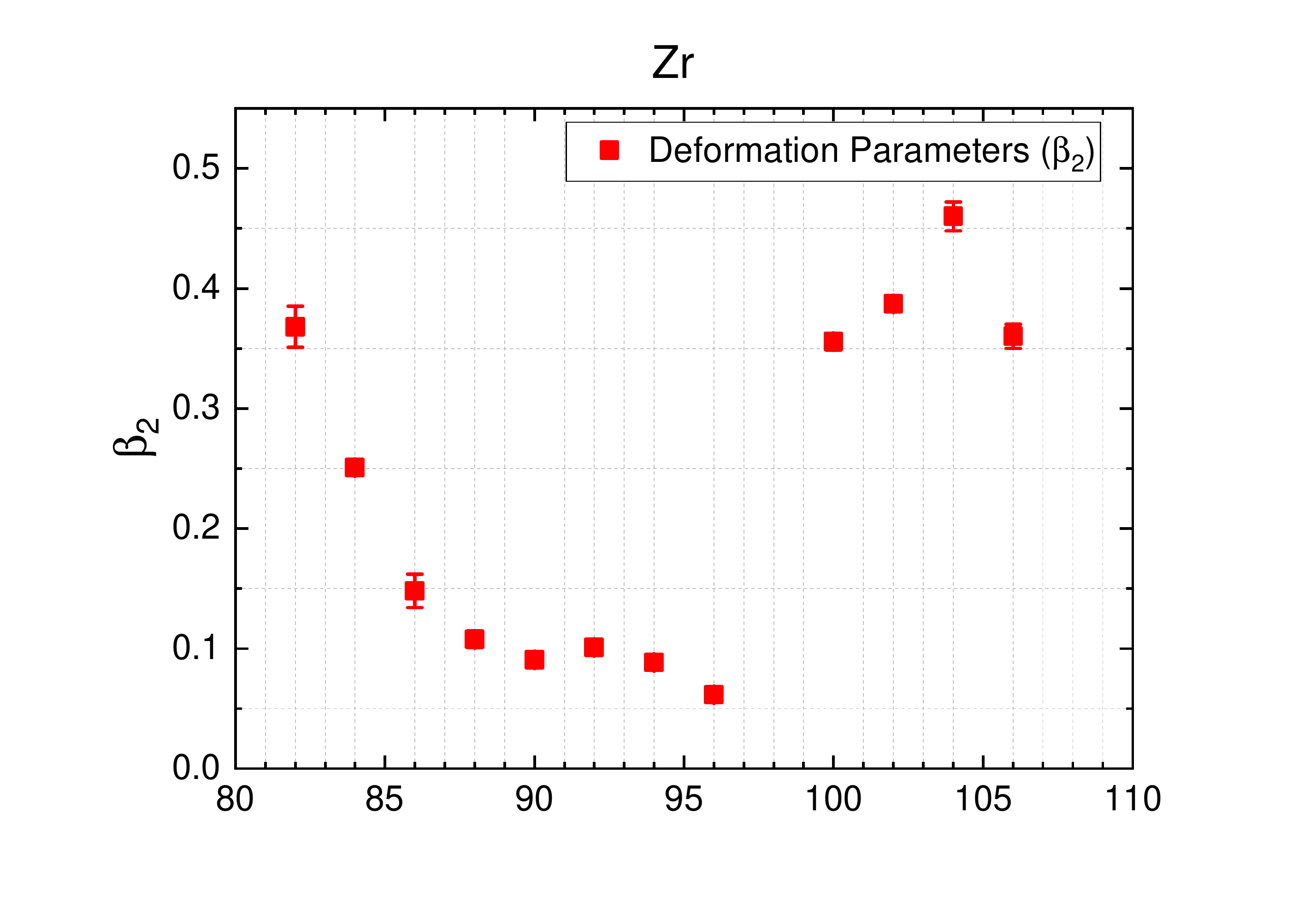}
  \end{minipage}%
  }\par
  \subfloat[]{%
  \begin{minipage}{\linewidth}
\centering
 \includegraphics[width=.6\linewidth]{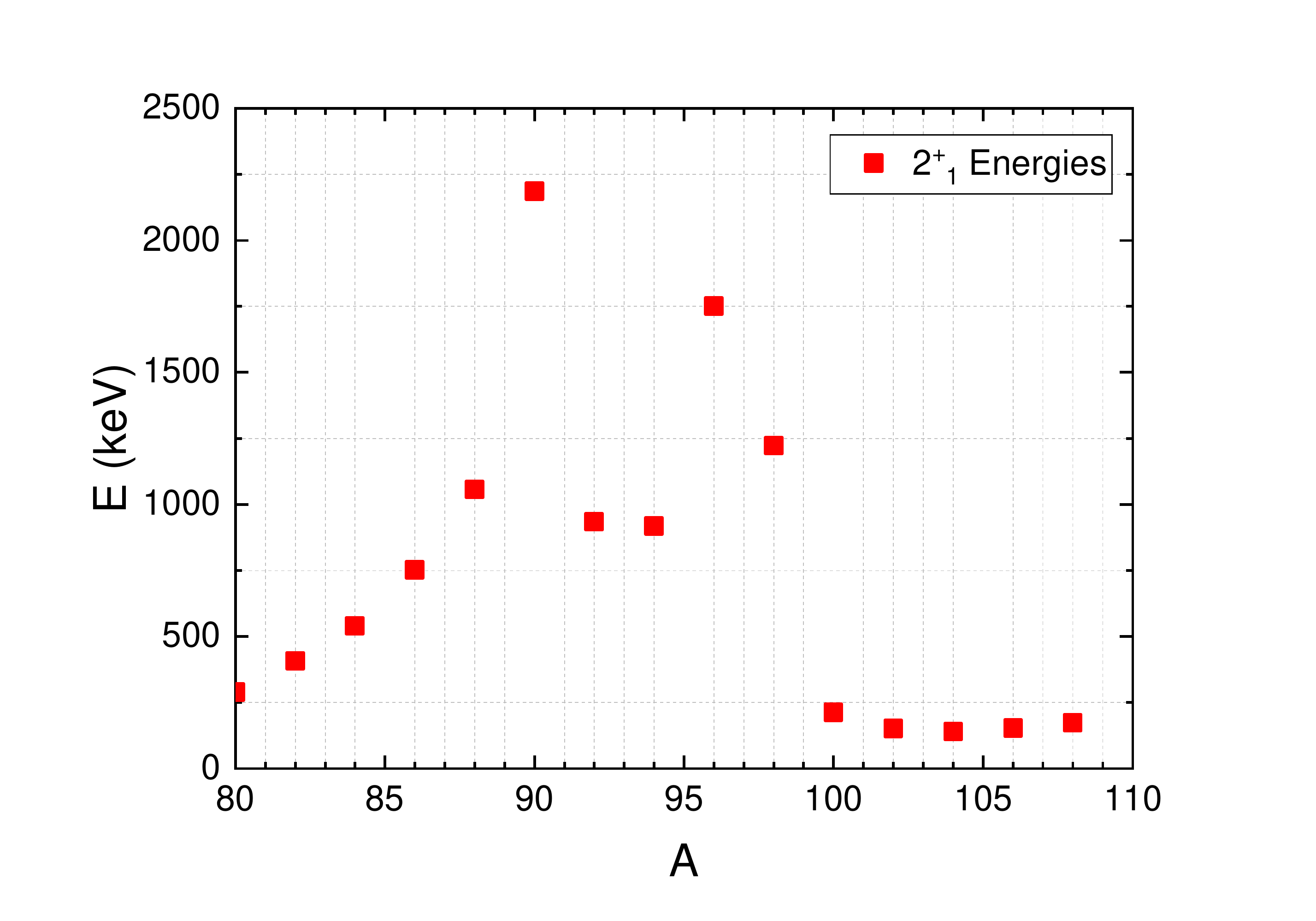}  \end{minipage}%
  }
  \caption{Quadrupole deformation ($\beta_2$) parameters (a), and the first 2$^+$ state energies (b)  in Zr. Data were taken from Ref.~\cite{Pri16}.}
\label{fig:beta2}
\end{figure}

Low-lying  0$^+$ states have been observed in many even-even nuclei. They alter the nuclear structure and decay properties  and impact basic and fundamental science applications. 
$^{96}$Zr is a long-lived radioactive isotope that can disintegrate via single and double-beta decay processes. It has many unique properties, including the lowest quadrupole deformation parameter ($\beta_2$) of   0.0615(33)~\cite{Pri16}  and the second-highest 2$^{+}_1 \rightarrow 0^{+}_1$  transition energy among zirconium nuclei.   
 Fig.~\ref{fig:beta2} data reveal $\beta_2$ value  in the $N=56$ $^{96}$Zr is surprisingly  lower than in the $N=50$ magic $^{90}$Zr. On the surface, it looks like a new ``magic" nucleus. At the same time, a recent attempt to calculate $^{96}$Zr double-beta decay half-life assuming the present value of quadrupole deformation overestimated experimental half-life by a factor of 80~\cite{Pri22}. 

\begin{figure}
  \subfloat[]{%
  \begin{minipage}{\linewidth}
\centering
 \includegraphics[width=.6\linewidth]{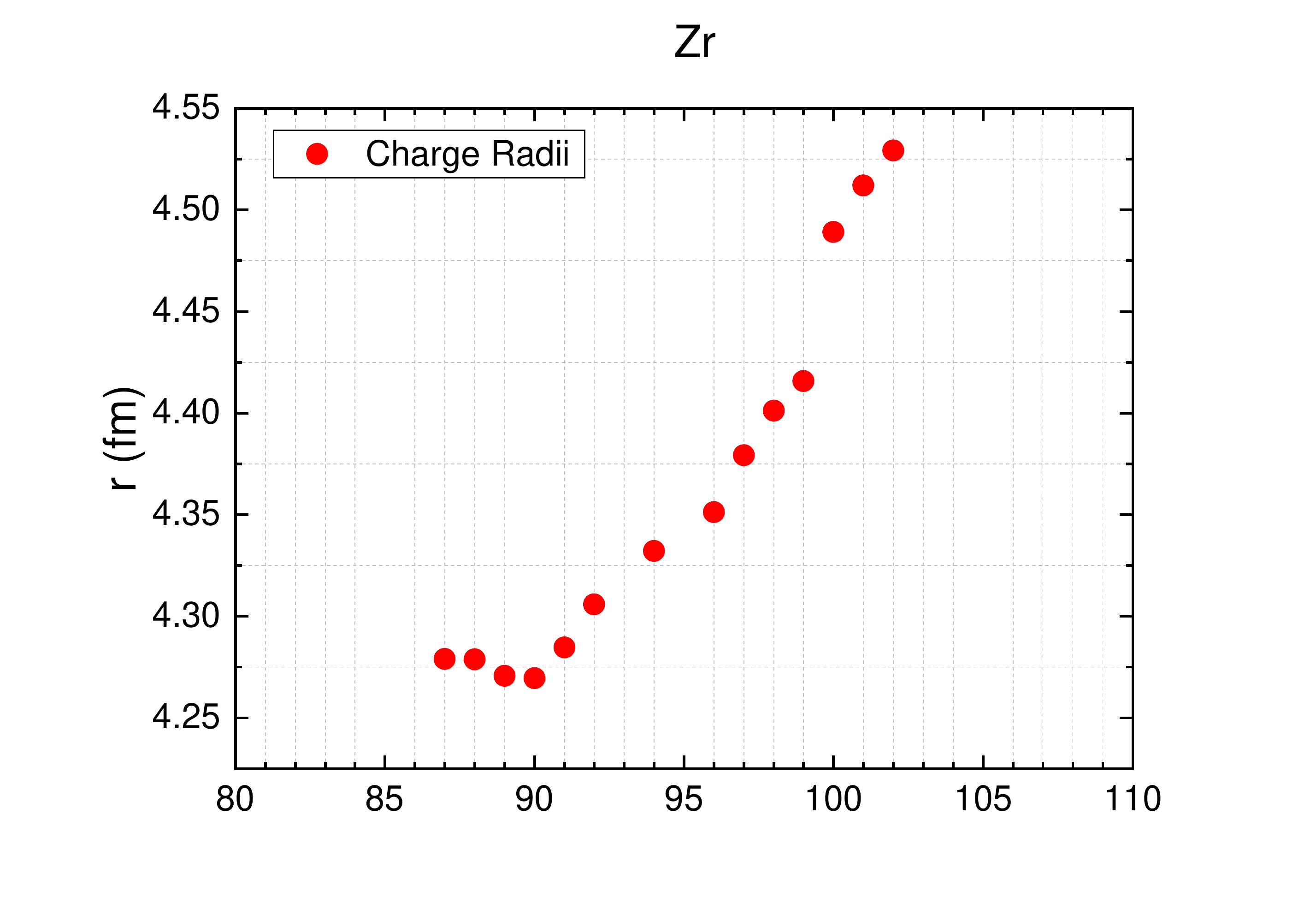}  \end{minipage}%
  }\par
  \subfloat[]{%
  \begin{minipage}{\linewidth}
\centering
   \includegraphics[width=.6\linewidth]{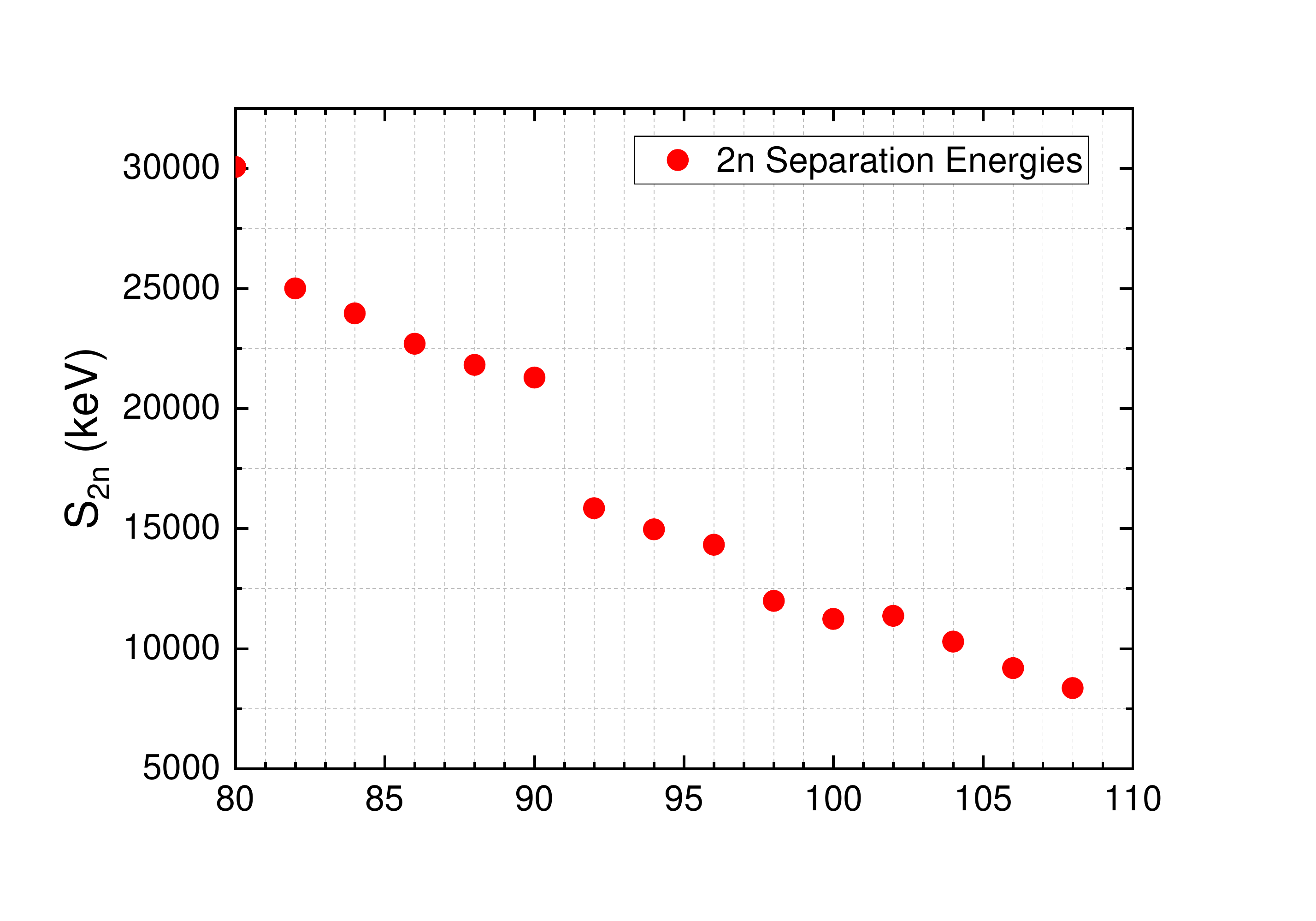}
     \end{minipage}%
  }\par
  \subfloat[]{%
  \begin{minipage}{\linewidth}
\centering
  \includegraphics[width=.6\linewidth]{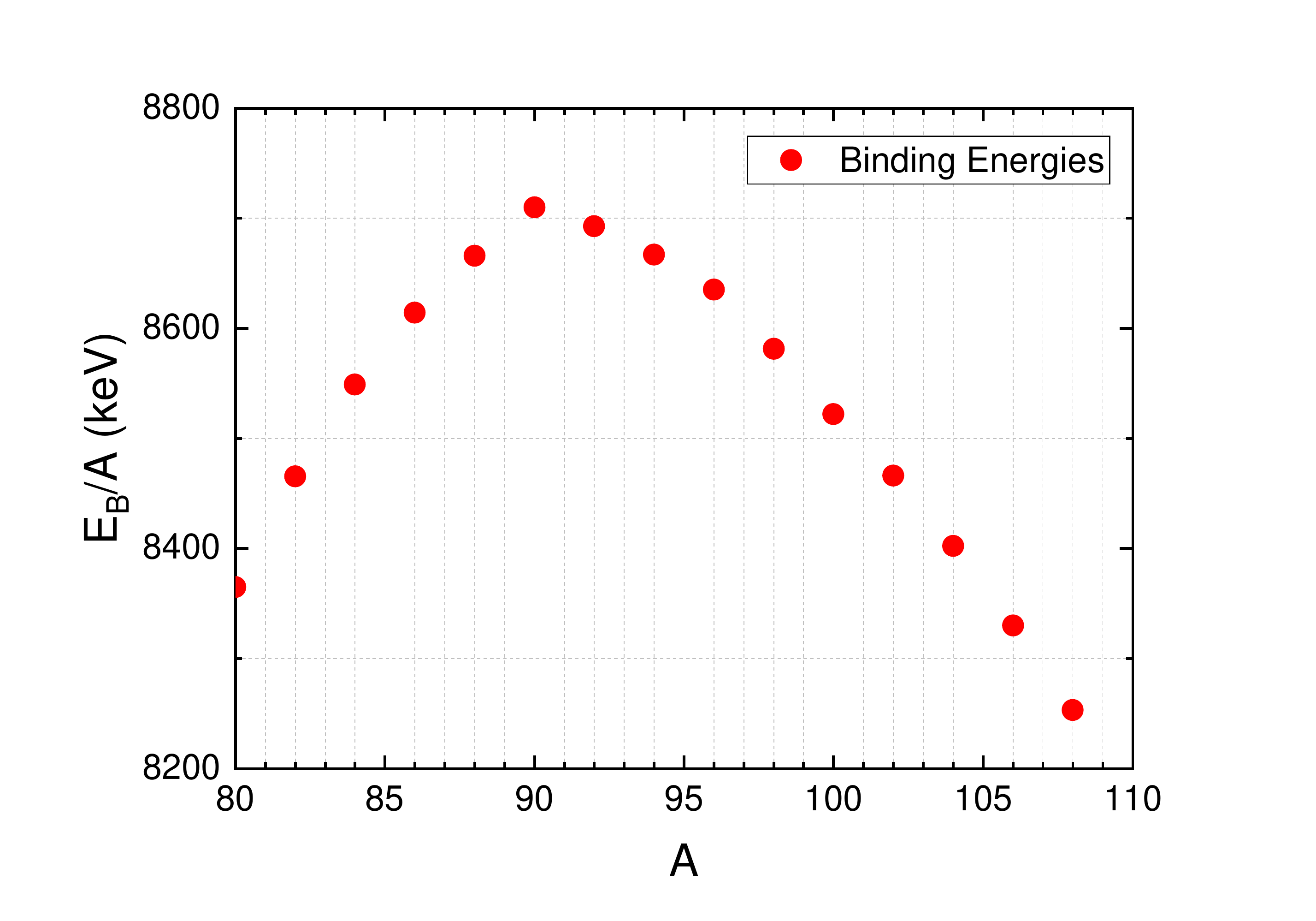}
    \end{minipage}%
  }
  \caption{Nuclear radii (a), 2n-separation energies (b), and nucleon binding energies (c) in Zr. Data were  taken from Ref.~\cite{Ang13,AME20}.}
\label{fig:shell}
\end{figure}

\begin{figure}
\centering
\includegraphics[width=0.8\textwidth]{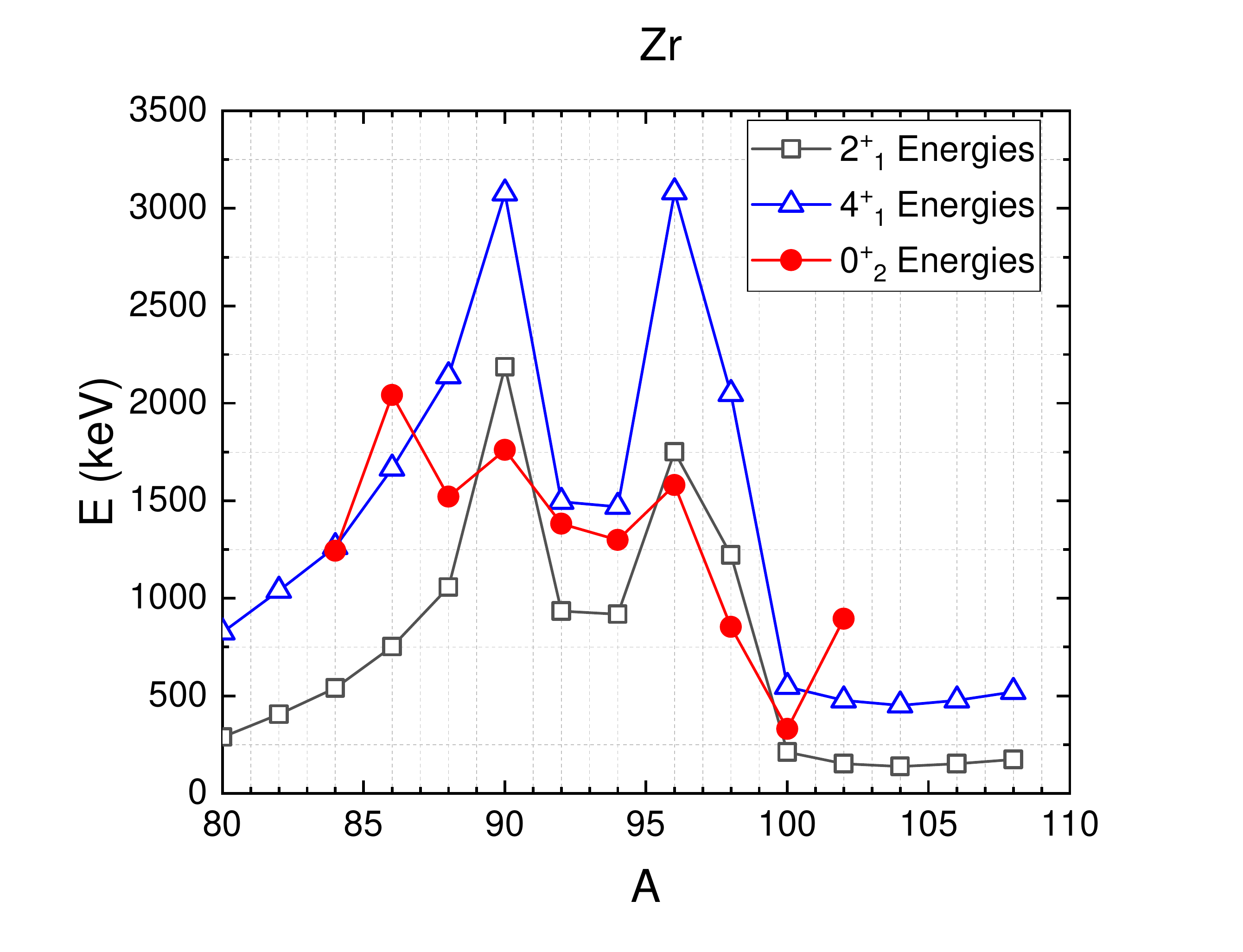}
\caption{Ground state band and 0$^+$ intruder states in Zr. Results are based on the ENSDF library~\cite{ensdf,Dim20}.}
\label{fig:levels}
\end{figure}

To investigate these magic-like properties, we would consider zirconium charge radii~\cite{Ang13}, two-neutron separation, and nucleon binding energies~\cite{AME20}. The Fig.~\ref{fig:shell} data on zirconium radii, 2n-separation, and binding energies indicate that $N=50$ is a good  magic number in zirconium, and we have to consider additional quantities for an understanding of the above-mentioned phenomena.
Complementary analysis of the ENSDF library~\cite{ensdf,Dim20} level schemes reveals intruder bands   in $^{90,96,98}$Zr as  shown in Fig.~\ref{fig:levels}. These bands are positioned below the 2$^+_1$ state energies, and the low-lying  0$^+_2$ state points to the shape coexistence phenomenon in zirconium nuclei. Further theoretical analysis~\cite{Gar19} shows that both spherically symmetrical and deformed states are present in  $^{90,96,98}$Zr rendering  $\beta_2$ values immaterial. This finding concurs with the Beliaev-Zelevinsky hypothesis~\cite{Bel62} that the formally-calculated $\beta_2$ noticeably smaller than 0.1 is too low for actual (not fluctuational) static deformation. The zirconium nuclei demonstrate the limits of applicability for deformation parameters  and the urgent need  for comprehensive analyses based on multiple quantities for the interpretation of nuclear physics observables. 

\subsection{1$^{-}_1$ and 3$^-_1$  and Other Excited States}

Low-lying 1$^-$ and 3$^-$ states are found in closed-shell carbon, oxygen, gadolinium, and lead nuclei. The nuclear shell model shows that the low-lying excited states in doubly-closed shell nuclei originate from (a) 1p-1h excitations that
can have a range of $J^{\pi}$ related to the  orbits near the Fermi surface and (b) 0$^{+}$ 2p-2h and 4p-4h "intruder" states. For $^{14}$O and $^{14}$C the lowest pair is p$_{1/2}$-s$_{1/2}$ leading to 0$^{-}$ and 1$^{-}$ states. 
Next to this p$_{1/2}$-d$_{5/2}$ configuration results in 2$^{-}$ and 3$^{-}$ states. All of these are low-lying excitations in $^{14}$0 and $^{14}$C, and the lowest state is defined by the Hamiltonian. 
For $^{146}$Gd and $^{208}$Pb, most of the lowest 1p-1h states change parity
and there are many ways to  assemble a given J. Thus it is natural that the lowest of these configurations should be the octupole collective 3$^{-}$ state.
These octupole states are experimentally observed, and the 3$^{-}_{1}$ excited-state in the doubly magic nucleus $^{208}$Pb is tied to a  large octupole collectivity of 34 Weisskopf units (W.u.)~\cite{Wol98,Rob16,Gaf12}. 

Lastly, the low-lying  1$^{-}_{1}$ states are found in $^{14}$C and $^{14}$O mirror nuclei with Z, N=8 shell closure. The Z, N=82 shell closures provide two cases of
the 3$^-_{1}$  states in $^{146}$Gd and $^{208}$Pb. Table~\ref{table:1-3-} data provide the full list and show the rationale for 1$^{-}_1$ and 3$^{-}_1$ spin and parity in even-even nuclei.

In $^{254}$Rf and $^{270}$Ds tentatively-assigned levels (8-) and (9-,10-), respectively, were reported~\cite{2015Da12,2015Ac04,2017Ac02}. As available experimental data for excited states for superheavy (Z$>$100) nuclei are 
generally rare, the listed excited states may or may not be the first excited states.

\subsection{2$^+_1$ Excited States}
Finally, ENSDF library numerical results for the lowest 2$^+$ excited states are shown in Table~\ref{table:2+}. 
The ENSDF library for individual nuclei is generally updated every 10 years or so, and it is educational to compare these data with a B(E2$\uparrow$) horizontal evaluation~\cite{Pri16} that was published six years ago. The comparison finds 10 first excited state half-lives that were introduced recently and absent in the horizontal evaluation. Supplemental analysis of the horizontal evaluation reveals 42  levels T$_{1/2}$ values that are missing in ENSDF. These latter values are included into Table~\ref{table:2+} for completeness. This combination of ENSDF and horizontal  B(E2$\uparrow$) evaluations produce the comprehensive up-to-date table for the 2$^+_{1}$ states.

\begin{center}
\begin{longtable}{l|ccccc}

\caption{List of 2$^+_1$  States in Even-Even Nuclei. Tentative spin and parity assignments are shown in parentheses,* or ** symbols for the nuclei where T$_{1/2}$ values  are  taken or updated, respectively,  using Ref.~\cite{Pri16}. $^{\dagger}$ symbol was used to mark nuclei where half-lives were recently updated in ENSDF, and missing in Ref.~\cite{Pri16}.}\\ 
\hline \hline
\textbf{Nuclide} & \textbf{Z} & \textbf{N} & \textbf{J$^{\pi}$} & \textbf{Energy, keV} & \textbf{T$_{1/2}$}  \\
\hline 
\endfirsthead
\hline
\textbf{Nuclide} & \textbf{Z} & \textbf{N} & \textbf{J$^{\pi}$} & \textbf{Energy, keV} & \textbf{T$_{1/2}$}  \\
\hline 
\endhead
\hline
\multicolumn{6}{r}{\textit{Continued on next page ...}} \\
\endfoot
\hline
\endlastfoot

$^{6}$He  & 2 & 4 & 2+ &  1797(25)   &  113(20) keV  \\
$^{8}$He  & 2 & 6 & 2+ &  3.10E+03(5)   &  0.6(2) MeV$^{\dagger}$  \\
$^{10}$He  & 2 & 8 &  (2+)   &  3.24E+03(20)   &  1000(300) keV$^{\dagger}$  \\
$^{6}$Be  & 4 & 2 &  (2+)  &  1670(50)   &  1.16(6) MeV$^{\dagger}$ \\
$^{8}$Be  & 4 & 4 & 2+ &  3030(10)   &  1513(15) keV $^{\dagger}$ \\
$^{10}$Be  & 4 & 6 & 2+ &  3368.03(3)   &  125(12) fs  \\
$^{12}$Be  & 4 & 8 & 2+ &  2109(1)   &  0.957(19) ps  \\
$^{10}$C  & 6 & 4 & 2+ &  3353.7(6)   &  107(17) fs  \\
$^{12}$C  & 6 & 6 & 2+ &  4439.82(21)   &  1.08E-02(6) eV  \\
$^{16}$C & 6 & 10 & 2+ &  1766(10)   &     9.2$^{+11}_{-9}$ ps$^*$  \\
$^{18}$C  & 6 & 12 & 2+ &  1588(8)   &  15.5(25) ps  \\
$^{20}$C  & 6 & 14 & 2+ &  1618(11)   &  6.8(20) ps  \\
$^{18}$O  & 8 & 10 & 2+ &  1982.07(9)   &  1.94(5) ps  \\
$^{20}$O  & 8 & 12 & 2+ &  1673.68(15)   &  7.3(3) ps  \\
$^{22}$O  & 8 & 14 & 2+ &  3199(8)   &  0.4(15) ps  \\
$^{24}$O  & 8 & 16 & 2+ &  4.79E+03(11)   &  0.05(16) MeV  \\
$^{26}$O  & 8 & 18 &  (2+)   &  1277(96)   &       \\
$^{16}$Ne  & 10 & 6 & 2+ &  1.77E+03(3)   &   $<$50   keV  \\
$^{18}$Ne  & 10 & 8 & 2+ &  1887.3(2)   &  0.46(4) ps  \\
$^{20}$Ne  & 10 & 10 & 2+ &  1633.674(15)   &  0.73(4) ps  \\
$^{22}$Ne  & 10 & 12 & 2+ &  1274.537(7)   &  3.6(5) ps  \\
$^{24}$Ne  & 10 & 14 & 2+ &  1981.6(4)   &  660(150) fs  \\
$^{26}$Ne  & 10 & 16 & 2+ &  2018(3)   &  0.6(8) ps  \\
$^{30}$Ne  & 10 & 20 &  (2+)   &  792(4)   &  20(28) ps  \\
$^{32}$Ne  & 10 & 22 &  (2+)   &  722(9)   &       \\
$^{20}$Mg  & 12 & 8 & 2+ &  1598(10)   &     1.36$^{+30}_{-21}$ ps$^*$   \\
$^{22}$Mg  & 12 & 10 & 2+ &  1247.02(3)   &  2.0(8) ps  \\
$^{24}$Mg  & 12 & 12 & 2+ &  1368.672(5)   &  1.33(6) ps  \\
$^{26}$Mg  & 12 & 14 & 2+ &  1808.74(4)   &  476(21) fs  \\
$^{28}$Mg  & 12 & 16 & 2+ &  1473.54(10)   &  1.2(1) ps  \\
$^{30}$Mg  & 12 & 18 & 2+ &  1482.8(3)   &  1.5(2) ps  \\
$^{32}$Mg  & 12 & 20 & 2+ &  885.3(1)   &  11.4(20) ps  \\
$^{34}$Mg  & 12 & 22 & 2+ &  660(7)   &  40(8) ps  \\
$^{36}$Mg & 12 & 24 &  (2+)   &  660(6)   &     43.6$^{+132}_{-90}$ ps$^*$  \\
$^{38}$Mg  & 12 & 26 &  (2+)   &  656(6)   &       \\
$^{24}$Si  & 14 & 10 & 2+ &  1879(11)   &  10(3) fs  \\
$^{26}$Si  & 14 & 12 & 2+ &  1797.3(10)   &  440(40) fs  \\
$^{28}$Si  & 14 & 14 & 2+ &  1779.03(11)   &  475(17) fs  \\
$^{30}$Si  & 14 & 16 & 2+ &  2235.322(18)   &  215(28) fs  \\
$^{32}$Si  & 14 & 18 & 2+ &  1941.4(3)   &  0.78(22) ps  \\
$^{34}$Si  & 14 & 20 & 2+ &  3327.14(20)   &  82(32) fs  \\
$^{36}$Si  & 14 & 22 & 2+ &  1408(10)   &  2.7(4) ps  \\
$^{38}$Si  & 14 & 24 & 2+ &  1074(2)   &  10(3) ps  \\
$^{40}$Si   & 14 & 26 &  (2+)   &  986(5)   &    7.1$^{+27}_{-15}$ ps$^*$   \\
$^{42}$Si  & 14 & 28 &  (2+)   &  742(8)   &       \\
$^{28}$S  & 16 & 12 & 2+ &  1507(7)   &  2.0(3) ps  \\
$^{30}$S  & 16 & 14 & 2+ &  2210.6(5)   &  156(9) fs  \\
$^{32}$S  & 16 & 16 & 2+ &  2230.57(15)   &  169(11) fs  \\
$^{34}$S  & 16 & 18 & 2+ &  2127.564(13)   &  318(8) fs  \\
$^{36}$S  & 16 & 20 & 2+ &  3290.9(3)   &  83(7) fs  \\
$^{38}$S  & 16 & 22 & 2+ &  1292.02(20)   &  3.3(1) ps  \\
$^{40}$S  & 16 & 24 & 2+ &  903.69(7)   &  14.1(3) ps  \\
$^{42}$S  & 16 & 26 & 2+ &  903(5)   &  11.9(20) ps  \\
$^{44}$S  & 16 & 28 & 2+ &  1329(5)   &  2.4(7) ps  \\
$^{46}$S  & 16 & 30 &  (2+)   &  952(8)   &       \\
$^{30}$Ar  & 18 & 12 &  (2+)   & 7.00E+02 &       \\
$^{32}$Ar  & 18 & 14 & 2+ &  1867(8)   &  0.46(12) ps  \\
$^{34}$Ar  & 18 & 16 & 2+ &  2091.1(3)   &  319(42) fs  \\
$^{36}$Ar  & 18 & 18 & 2+ &  1970.38(5)   &  328(20) fs  \\
$^{38}$Ar  & 18 & 20 & 2+ &  2167.472(12)   &  0.458(21) ps  \\
$^{40}$Ar  & 18 & 22 & 2+ &  1460.849(5)   &  1.15(5) ps  \\
$^{42}$Ar  & 18 & 24 & 2+ &  1208.22(13)   &  2.6(1) ps  \\
$^{46}$Ar  & 18 & 28 & 2+ &  1577(1)   &  1.59(32) ps  \\
$^{48}$Ar  & 18 & 30 &  (2+)   &  1038(6)   &     2.1$^{+4}_{-3}$ ps$^*$      \\
$^{50}$Ar  & 18 & 32 &  (2+)   &  1178(18)   &       \\
$^{36}$Ca  & 20 & 16 &  (2+)   &  3045(24)   &       \\
$^{38}$Ca  & 20 & 18 & 2+ &  2213.2(10)   &  0.56(6) ps  \\
$^{42}$Ca  & 20 & 22 & 2+ &  1524.71(3)   &  0.83(3) ps  \\
$^{44}$Ca  & 20 & 24 & 2+ &  1157.019(4)   &  2.71(15) ps  \\
$^{46}$Ca  & 20 & 26 & 2+ &  1346(3)   &  3.6(3) ps  \\
$^{48}$Ca  & 20 & 28 & 2+ &  3831.72(6)   &  38.7(19) fs  \\
$^{50}$Ca  & 20 & 30 & 2+ &  1026.72(10)   &  66.5(21) ps  \\
$^{52}$Ca  & 20 & 32 & 2+ &  2563(1)   &       \\
$^{54}$Ca  & 20 & 34 &  (2+)   &  2043(19)   &       \\
$^{42}$Ti  & 22 & 20 & 2+ &  1554.6(3)   &  0.44(11) ps  \\
$^{44}$Ti  & 22 & 22 & 2+ &  1083.06(9)   &  3.1(8) ps  \\
$^{46}$Ti  & 22 & 24 & 2+ &  889.286(3)   &  5.32(15) ps  \\
$^{48}$Ti  & 22 & 26 & 2+ &  983.539(24)   &  4.04(10) ps  \\
$^{50}$Ti  & 22 & 28 & 2+ &  1553.794(8)   &  1.047(35) ps  \\
$^{52}$Ti  & 22 & 30 & 2+ &  1050.06(9)   &  3.6(14) ps  \\
$^{54}$Ti  & 22 & 32 &  (2+)   &  1494.8(8)   &  1.06(19) ps  \\
$^{56}$Ti    & 22 & 34 &  (2+)   &  1128.2(4)   &    2.6$^{+13}_{-6}$ ps$^*$    \\
$^{58}$Ti & 22 & 36 & 2+ &  1047(4)   &     5.4$^{+65}_{-21}$ ps$^*$   \\
$^{60}$Ti  & 22 & 38 &  (2+)   &  850(5)   &       \\
$^{46}$Cr  & 24 & 22 & 2+ &  892.16(10)   &  5.4(12) ps  \\
$^{48}$Cr  & 24 & 24 & 2+ &  752.19(11)   &  7.3(8) ps  \\
$^{50}$Cr  & 24 & 26 & 2+ &  783.31(10)   &  9.08(28) ps  \\
$^{52}$Cr  & 24 & 28 & 2+ &  1434.091(14)   &  0.783(21) ps  \\
$^{54}$Cr  & 24 & 30 & 2+ &  834.855(3)   &  8(3) ps  \\
$^{56}$Cr  & 24 & 32 & 2+ &  1006.61(20)   &  $\geq$1.4   ps  \\
$^{58}$Cr  & 24 & 34 & 2+ &  880.7(2)   &  5.4(9) ps  \\
$^{60}$Cr  & 24 & 36 &  (2+)   &  643.9(20)   &  23(3) ps  \\
$^{62}$Cr   & 24 & 38 &  (2+)   &  446(1)   &     91.5$^{+76}_{-90}$ ps$^*$   \\
$^{64}$Cr  & 24 & 40 & 2+ &  430(2)   &  123(19) ps  \\
$^{66}$Cr  & 24 & 42 &  (2+)   &  386(10)   &       \\
$^{50}$Fe  & 26 & 24 & 2+ &  764.9(3)   &  7.7(17) ps  \\
$^{52}$Fe  & 26 & 26 & 2+ &  849.45(10)   &  7.8(10) ps  \\
$^{54}$Fe  & 26 & 28 & 2+ &  1408.19(19)   &  0.76(2) ps  \\
$^{56}$Fe  & 26 & 30 & 2+ &  846.7778(19)   &  6.07(23) ps  \\
$^{58}$Fe  & 26 & 32 & 2+ &  810.7662(20)   &  6.54(19) ps  \\
$^{60}$Fe  & 26 & 34 & 2+ &  823.83(9)   &  7.9(8) ps  \\
$^{62}$Fe  & 26 & 36 & 2+ &  877.31(10)   &  5.3(6) ps  \\
$^{64}$Fe   & 26 & 38 & 2+ &  746.4(10)   &  7.1$^{+4}_{-8}$ ps$^*$     \\
$^{66}$Fe & 26 & 40 &  (2+)   &  574.4(10)   &   29.7$^{+21}_{-18}$ ps$^*$    \\
$^{68}$Fe  & 26 & 42 &  (2+)   &  522(10)   &   43.0$^{+60}_{-46}$ ps$^*$     \\
$^{70}$Fe  & 26 & 44 &  (2+)   &  480(13)   &       \\
$^{52}$Ni  & 28 & 24 & 2+ &  1397(6)   &       \\
$^{54}$Ni  & 28 & 26 & 2+ &  1392.3(4)   &  0.89(17) ps  \\
$^{56}$Ni  & 28 & 28 & 2+ &  2700.6(7)   &  53(17) fs  \\
$^{58}$Ni  & 28 & 30 & 2+ &  1454.21(9)   &  0.652(21) ps  \\
$^{60}$Ni  & 28 & 32 & 2+ &  1332.514(4)   &  0.735(21) ps  \\
$^{62}$Ni  & 28 & 34 & 2+ &  1172.98(10)   &  1.45(4) ps  \\
$^{64}$Ni  & 28 & 36 & 2+ &  1345.75(5)   &  1.088(35) ps  \\
$^{66}$Ni  & 28 & 38 & 2+ &  1424.8(10)   &  0.8(2) ps  \\
$^{70}$Ni  & 28 & 42 & 2+ &  1259.55(5)   &  1.04(17) ps  \\
$^{72}$Ni  & 28 & 44 &  (2+)   &  1096(20)   &       \\
$^{74}$Ni  & 28 & 46 & 2+ &  1024(1)   &  3.9(11) ps  \\
$^{76}$Ni  & 28 & 48 &  (2+)   &  992(2)   &       \\
$^{58}$Zn  & 30 & 28 &  (2+)   &  1356(3)   &       \\
$^{60}$Zn  & 30 & 30 & 2+ &  1003.9(20)   &       \\
$^{62}$Zn  & 30 & 32 & 2+ &  953.84(9)   &  2.93(14) ps  \\
$^{64}$Zn  & 30 & 34 & 2+ &  991.56(5)   &  1.94(5) ps  \\
$^{66}$Zn  & 30 & 36 & 2+ &  1039.2279(21)   &  1.68(3) ps  \\
$^{68}$Zn  & 30 & 38 & 2+ &  1077.37(4)   &  1.61(2) ps  \\
$^{70}$Zn  & 30 & 40 & 2+ &  884.92(8)   &  3.65(21) ps  \\
$^{72}$Zn  & 30 & 42 & 2+ &  652.7(5)   &  14(3) ps  \\
$^{74}$Zn  & 30 & 44 & 2+ &  605.9(8)   &  17.7(13) ps  \\
$^{76}$Zn  & 30 & 46 &  (2+)   &  598.68(10)   &   25.4$^{+31}_{-31}$ ps$^*$     \\
$^{78}$Zn  & 30 & 48 & 2+ &  730.2(4)   &  18(4) ps  \\
$^{80}$Zn  & 30 & 50 & 2+ &  1492(1)   &  0.52(11) ps  \\
$^{82}$Zn  & 30 & 52 &  (2+)   &  618(15)   &       \\
$^{64}$Ge  & 32 & 32 & 2+ &  901.7(3)   &    2.3$^{+3}_{-3}$ ps$^*$     \\
$^{66}$Ge  & 32 & 34 & 2+ &  956.94(8)   &  3.7(7) ps  \\
$^{68}$Ge  & 32 & 36 & 2+ &  1015.81(8)   &  2.08(11) ps  \\
$^{70}$Ge  & 32 & 38 & 2+ &  1039.506(9)   &  1.31(2) ps  \\
$^{74}$Ge  & 32 & 42 & 2+ &  595.85(6)   &  12.41(9) ps  \\
$^{76}$Ge  & 32 & 44 & 2+ &  562.93(3)   &  18.2(2) ps  \\
$^{78}$Ge  & 32 & 46 & 2+ &  619.36(12)   &  13.5(24) ps  \\
$^{80}$Ge  & 32 & 48 & 2+ &  659.15(4)   &  16.4(32) ps  \\
$^{82}$Ge  & 32 & 50 & 2+ &  1348.3(1)   &  0.5(8) ps  \\
$^{84}$Ge  & 32 & 52 &  (2+)   &  624.3(7)   &       \\
$^{86}$Ge  & 32 & 54 &  (2+)   & 527 &       \\
$^{66}$Se  & 34 & 32 &  (2+)   &  929(7)   &       \\
$^{68}$Se  & 34 & 34 & 2+ &  853.75(21)   &  2.8(4) ps  \\
$^{70}$Se  & 34 & 36 & 2+ &  944.52(5)   &  2.23(14) ps  \\
$^{72}$Se  & 34 & 38 & 2+ &  862.07(8)   &  2.82(20) ps  \\
$^{74}$Se  & 34 & 40 & 2+ &  634.74(6)   &  7.08(9) ps  \\
$^{76}$Se  & 34 & 42 & 2+ &  559.102(5)   &  12.3(2) ps  \\
$^{78}$Se  & 34 & 44 & 2+ &  613.727(3)   &  9.79(21) ps  \\
$^{80}$Se  & 34 & 46 & 2+ &  666.27(7)   &  8.52(21) ps  \\
$^{82}$Se  & 34 & 48 & 2+ &  654.71(16)   &  12.8(7) ps  \\
$^{84}$Se  & 34 & 50 & 2+ &  1454.55(8)   &  0.42(7) ps  \\
$^{86}$Se  & 34 & 52 & 2+ &  704.3(5)   &  7.5(22) ps  \\
$^{90}$Se  & 34 & 56 &  (2+)   &  547(8)   &       \\
$^{74}$Kr  & 36 & 38 & 2+ &  455.61(10)   &  23.4(4) ps  \\
$^{76}$Kr  & 36 & 40 & 2+ &  423.96(7)   &  24.9(7) ps  \\
$^{78}$Kr  & 36 & 42 & 2+ &  455.033(23)   &  21.6(7) ps  \\
$^{80}$Kr  & 36 & 44 & 2+ &  616.6(10)   &  8.3(5) ps  \\
$^{82}$Kr  & 36 & 46 & 2+ &  776.526(8)   &  4.45(18) ps  \\
$^{84}$Kr  & 36 & 48 & 2+ &  881.615(3)   &  4.05(13) ps  \\
$^{86}$Kr  & 36 & 50 & 2+ &  1564.61(7)   &  0.286(4) ps  \\
$^{88}$Kr  & 36 & 52 & 2+ &  775.32(4)   &  11.1(12) ps  \\
$^{90}$Kr  & 36 & 54 & 2+ &  707.12(5)   &  10.7(16) ps  \\
$^{92}$Kr  & 36 & 56 & 2+ &  769.1(5)   &   5.1$^{+18}_{-16}$ ps$^{**}$ \\
$^{94}$Kr   & 36 & 58 & 2+ & 665.5 &    8.7$^{+11}_{-9}$ ps$^*$   \\
$^{96}$Kr  & 36 & 60 &  (2+)   &  554.1(5)   &  12.4(8) ps  \\
$^{98}$Kr  & 36 & 62 &  (2+)   &  329(7)   &       \\
$^{100}$Kr  & 36 & 64 &  (2+)   &  309(10)   &       \\
$^{74}$Sr  & 38 & 36 &  (2+)   &  471(1)   &       \\
$^{76}$Sr & 38 & 38 &  (2+)   &  262.3(2)   &     204.4$^{+256}_{-256}$ ps$^*$    \\
$^{78}$Sr  & 38 & 40 & 2+ &  277.6(10)   &  155(19) ps  \\
$^{80}$Sr  & 38 & 42 & 2+ &  385.88(8)   &  34.2(12) ps  \\
$^{82}$Sr  & 38 & 44 & 2+ &  573.54(8)   &  8.9(4) ps  \\
$^{84}$Sr  & 38 & 46 & 2+ &  793.22(6)   &  3.23(35) ps  \\
$^{86}$Sr  & 38 & 48 & 2+ &  1076.68(4)   &  1.46(1) ps  \\
$^{88}$Sr  & 38 & 50 & 2+ &  1836.09(8)   &  0.154(8) ps  \\
$^{90}$Sr  & 38 & 52 & 2+ &  831.68(4)   &  7(2) ps  \\
$^{92}$Sr  & 38 & 54 & 2+ &  814.98(3)   &  8(3) ps  \\
$^{94}$Sr  & 38 & 56 & 2+ &  836.9(1)   &  6.9(28) ps  \\
$^{96}$Sr  & 38 & 58 & 2+ &  814.93(7)   &  4.8(28) ps  \\
$^{98}$Sr  & 38 & 60 & 2+ &  144.7(5)   &  2.78(8) ns  \\
$^{100}$Sr  & 38 & 62 &  (2+)   &  129.18(9)   &  3.91(16) ns  \\
$^{102}$Sr  & 38 & 64 &  (2+)   &  126(2)   &  3.0(12) ns  \\
$^{80}$Zr  & 40 & 40 &  (2+)   &  288.9(2)   &       \\
$^{82}$Zr  & 40 & 42 & 2+ &  407(10)   &  22(9) ps  \\
$^{84}$Zr  & 40 & 44 & 2+ &  539.92(9)   &  14.1(8) ps  \\
$^{86}$Zr  & 40 & 46 & 2+ &  751.75(3)   &  7.5(14) ps  \\
$^{88}$Zr  & 40 & 48 & 2+ &  1057.03(4)   &  2.5(28) ps  \\
$^{92}$Zr  & 40 & 52 & 2+ &  934.51(4)   &  5.0(4) ps  \\
$^{94}$Zr  & 40 & 54 & 2+ &  918.75(5)   &  6.9(15) ps  \\
$^{100}$Zr  & 40 & 60 & 2+ &  212.61(4)   &  0.574(15) ns  \\
$^{102}$Zr  & 40 & 62 & 2+ &  151.78(11)   &  1.8(4) ns  \\
$^{104}$Zr  & 40 & 64 &  (2+)   &  139.3(3)   &  2.0(3) ns  \\
$^{106}$Zr  & 40 & 66 &  (2+)   &  152.1(5)   & 1802$^{+139}_{-104}$ ps$^*$     \\
$^{108}$Zr  & 40 & 68 &  (2+)   &  174.3(5)   &       \\
$^{84}$Mo  & 42 & 42 &  (2+)   &  443.9(2)   &       \\
$^{86}$Mo  & 42 & 44 &  (2+)   &  566.6(4)   &       \\
$^{88}$Mo  & 42 & 46 & 2+ &  740.54(4)   &  7.14(21) ps$^{\dagger}$   \\
$^{90}$Mo  & 42 & 48 & 2+ &  948.02(9)   &       \\
$^{92}$Mo  & 42 & 50 & 2+ &  1509.51(3)   &  0.35(2) ps  \\
$^{94}$Mo  & 42 & 52 & 2+ &  871.098(16)   &  2.77(6) ps  \\
$^{96}$Mo  & 42 & 54 & 2+ &  778.237(10)   &  3.67(6) ps  \\
$^{100}$Mo  & 42 & 58 & 2+ &  535.59(4)   &  12.4(3) ps  \\
$^{102}$Mo  & 42 & 60 & 2+ &  296.61(4)   &  125(4) ps  \\
$^{104}$Mo  & 42 & 62 & 2+ &  192.19(9)   &  0.97(8) ns  \\
$^{106}$Mo  & 42 & 64 & 2+ &  171.549(8)   &  1.25(3) ns  \\
$^{108}$Mo  & 42 & 66 & 2+ &  192.79(15)   &  0.5(3) ns  \\
$^{110}$Mo  & 42 & 68 &  (2+)   &  213.77(10)   &       \\
$^{88}$Ru  & 44 & 44 &  (2+)   &  616.2(5)   &       \\
$^{90}$Ru  & 44 & 46 & 2+ &  738.1(10)   &       \\
$^{92}$Ru  & 44 & 48 &  (2+)   &  865.7(1)   &       \\
$^{94}$Ru  & 44 & 50 & 2+ &  1430.71(20)   &       \\
$^{96}$Ru  & 44 & 52 & 2+ &  832.56(5)   &  2.94(6) ps  \\
$^{98}$Ru  & 44 & 54 & 2+ &  652.46(5)   &  5.96(20) ps  \\
$^{100}$Ru  & 44 & 56 & 2+ &  539.5103(20)   &  12.54(10) ps  \\
$^{102}$Ru  & 44 & 58 & 2+ &  475.0962(10)   &  18.4(3) ps  \\
$^{104}$Ru  & 44 & 60 & 2+ &  358.02(7)   &  56.4(10) ps  \\
$^{106}$Ru  & 44 & 62 & 2+ &  270.07(4)   &  0.20(3) ns  \\
$^{108}$Ru  & 44 & 64 & 2+ &  242.23(4)   &  0.36(3) ns  \\
$^{110}$Ru  & 44 & 66 & 2+ &  240.73(8)   &  0.32(2) ns  \\
$^{112}$Ru  & 44 & 68 & 2+ &  236.69(16)   &  0.32(3) ns  \\
$^{114}$Ru  & 44 & 70 & 2+ &  265.19(17)   &       \\
$^{116}$Ru  & 44 & 72 &  (2+)   &  292.43(21)   &       \\
$^{118}$Ru  & 44 & 74 &  (2+)   &  327.6(3)   &       \\
$^{92}$92Pd  & 46 & 46 &  (2+)   &  873.6(2)   &       \\
$^{94}$94Pd  & 46 & 48 & 2+ &  813.8(10)   &       \\
$^{96}$96Pd  & 46 & 50 & 2+ &  1415.31(10)   &       \\
$^{98}$98Pd  & 46 & 52 & 2+ &  862.69(11)   &  $<$11.3   ps  \\
$^{100}$Pd  & 46 & 54 & 2+ &  665.49(10)   &  6.24(28) ps  \\
$^{102}$Pd  & 46 & 56 & 2+ &  556.44(5)   &  11.5(8) ps  \\
$^{104}$Pd  & 46 & 58 & 2+ &  555.81(4)   &  9.9(5) ps  \\
$^{106}$Pd  & 46 & 60 & 2+ &  511.85(23)   &  12.2(4) ps  \\
$^{108}$Pd  & 46 & 62 & 2+ &  433.938(4)   &  23.9(7) ps  \\
$^{110}$Pd  & 46 & 64 & 2+ &  373.8(7)   &  44(7) ps  \\
$^{112}$Pd  & 46 & 66 & 2+ &  348.66(13)   &  84(14) ps  \\
$^{114}$Pd  & 46 & 68 & 2+ &  332.61(10)   &  82(14) ps  \\
$^{116}$Pd  & 46 & 70 & 2+ &  340.26(8)   &  0.11(3) ns  \\
$^{118}$Pd  & 46 & 72 &  (2+)   &  378.6(1)   &       \\
$^{120}$Pd  & 46 & 74 &  (2+)   &  438(10)   &       \\
$^{122}$Pd  & 46 & 76 &  (2+)   &  499(9)   &       \\
$^{124}$Pd  & 46 & 78 &  (2+)   &  590(11)   &       \\
$^{126}$Pd  & 46 & 80 &  (2+)   &  693.3(5)   &       \\
$^{128}$Pd  & 46 & 82 &  (2+)   &  1311.4(5)   &       \\
$^{98}$Cd  & 48 & 50 &  (2+)   &  1395.1(2)   &       \\
$^{100}$Cd  & 48 & 52 & 2+ &  1004.11(10)   &  $>$1.5   ps  \\
$^{102}$Cd  & 48 & 54 & 2+ &  776.55(14)   &  3.5(6) ps  \\
$^{104}$Cd  & 48 & 56 & 2+ &  658(20)   &  6.3(21) ps  \\
$^{106}$Cd  & 48 & 58 & 2+ &  632.64(4)   &  7.27(8) ps  \\
$^{108}$Cd  & 48 & 60 & 2+ &  632.988(15)   &  6.86(7) ps  \\
$^{110}$Cd  & 48 & 62 & 2+ &  657.7623(11)   &  5.42(16) ps  \\
$^{112}$Cd  & 48 & 64 & 2+ &  617.518(3)   &  6.46(4) ps  \\
$^{114}$Cd  & 48 & 66 & 2+ &  558.456(2)   &  10.2(6) ps  \\
$^{116}$Cd  & 48 & 68 & 2+ &  513.49(15)   &  14.1(5) ps  \\
$^{118}$Cd  & 48 & 70 & 2+ &  487.77(8)   &  17.9(15) ps  \\
$^{120}$Cd  & 48 & 72 & 2+ &  505.94(17)   &  18(21) ps  \\
$^{122}$Cd  & 48 & 74 & 2+ &  569.45(8)   &  10(5) ps  \\
$^{124}$Cd   & 48 & 76 &  (2+)   &  612.8(4)   &  9.3$^{+6}_{-5}$ ps$^*$      \\
$^{126}$Cd   & 48 & 78 &  (2+)   &  652(9)   &   9.1$^{+27}_{-17}$ ps$^*$      \\
$^{128}$Cd  & 48 & 80 &  (2+)   &  645.8(20)   &       \\
$^{130}$Cd  & 48 & 82 &  (2+)   &  1325(1)   &       \\
$^{132}$Cd  & 48 & 84 &  (2+)   &  618(8)   &       \\
$^{104}$Sn  & 50 & 54 & 2+ &  1260.1(3)   &   0.51$^{+10}_{-7}$ ps$^*$     \\
$^{106}$Sn   & 50 & 56 & 2+ &  1207.7(5)   &    0.53$^{+8}_{-8}$ ps$^*$      \\
$^{108}$Sn  & 50 & 58 & 2+ &  1206.07(10)   &  0.48(12) ps  \\
$^{110}$Sn  & 50 & 60 & 2+ &  1212.02(9)   &  0.48(4) ps  \\
$^{112}$Sn  & 50 & 62 & 2+ &  1256.69(4)   &  0.376(5) ps  \\
$^{114}$Sn  & 50 & 64 & 2+ &  1299.907(7)   &  0.42(3) ps  \\
$^{116}$Sn  & 50 & 66 & 2+ &  1293.56(8)   &  0.374(10) ps  \\
$^{118}$Sn  & 50 & 68 & 2+ &  1229.666(16)   &  0.485(19) ps  \\
$^{120}$Sn  & 50 & 70 & 2+ &  1171.265(15)   &  0.64(12) ps  \\
$^{122}$Sn  & 50 & 72 & 2+ &  1140.51(3)   &  0.776(16) ps  \\
$^{124}$Sn  & 50 & 74 & 2+ &  1131.739(17)   &  0.92(3) ps  \\
$^{126}$Sn   & 50 & 76 & 2+ &  1141.15(4)   &   1.15$^{+7}_{-7}$ ps$^*$      \\
$^{128}$Sn  & 50 & 78 &  (2+)  &  1168.82(4)   &  1.63(10) ps  \\
$^{130}$Sn  & 50 & 80 &  (2+)   &  1221.26(5)   &     4.50$^{+97}_{-97}$ ps$^*$     \\
$^{132}$Sn  & 50 & 82 & 2+ &  4041.2(15)   &  2.4(4) fs  \\
$^{134}$Sn  & 50 & 84 & 2+ & 725.6 &     48.4$^{+100}_{-71}$ ps$^*$   \\
$^{136}$Sn  & 50 & 86 &  (2+)   &  688(1)   &       \\
$^{138}$Sn  & 50 & 88 &  (2+)   &  715(1)   &       \\
$^{106}$Te  & 52 & 54 &  (2+)   &  664.8(3)   &       \\
$^{108}$Te   & 52 & 56 & 2+ &  625.2(20)   &  7.6$^{+9}_{-9}$ ps$^*$    \\
$^{110}$Te  & 52 & 58 & 2+ &  657.2(3)   &       \\
$^{112}$Te & 52 & 60 & 2+ &  689(20)   &  3.95$^{+35}_{-35}$ ps$^*$      \\
$^{114}$Te  & 52 & 62 & 2+ &  708.74(15)   &  2.83(23) ps  \\
$^{116}$Te  & 52 & 64 & 2+ &  678.92(3)   &       \\
$^{118}$Te  & 52 & 66 & 2+ &  605.706(20)   &   6.1$^{+10}_{-10}$ ps$^*$     \\
$^{120}$Te  & 52 & 68 & 2+ &  560.438(20)   &  9.3(19) ps  \\
$^{122}$Te  & 52 & 70 & 2+ &  564.094(16)   &  7.46(5) ps  \\
$^{124}$Te  & 52 & 72 & 2+ &  602.7271(21)   &  6.2(1) ps  \\
$^{126}$Te  & 52 & 74 & 2+ &  666.352(10)   &  4.52(10) ps  \\
$^{128}$Te  & 52 & 76 & 2+ &  743.216(17)   &  3.3(3) ps  \\
$^{130}$Te  & 52 & 78 & 2+ &  839.494(17)   &  2.3(5) ps  \\
$^{132}$Te  & 52 & 80 & 2+ &  974.22(9)   &  1.83(18) ps  \\
$^{134}$Te  & 52 & 82 & 2+ &  1279.11(10)   &  0.64(20) ps  \\
$^{136}$Te  & 52 & 84 & 2+ &  606.64(5)   &  21.6(41) ps  \\
$^{138}$Te  & 52 & 86 &  (2+)   &  460.8(5)   &       \\
$^{140}$Te  & 52 & 88 &  (2+)   &  422.9(3)   &       \\
$^{110}$Xe  & 54 & 56 &  (2+)   &  469.7(20)   &       \\
$^{112}$Xe  & 54 & 58 & 2+ &  466(20)   &       \\
$^{114}$Xe  & 54 & 60 & 2+ &  450.08(19)   &  15.6(8) ps  \\
$^{116}$Xe  & 54 & 62 & 2+ &  393.6(2)   &  24.3(9) ps  \\
$^{118}$Xe  & 54 & 64 & 2+ &  337.32(13)   &  45(2) ps  \\
$^{120}$Xe  & 54 & 66 & 2+ &  322.61(4)   &  45.7(20) ps  \\
$^{122}$Xe  & 54 & 68 & 2+ &  331.28(7)   &  49.3(20) ps  \\
$^{124}$Xe  & 54 & 70 & 2+ &  354.03(4)   &  46.8(12) ps  \\
$^{126}$Xe  & 54 & 72 & 2+ &  388.631(9)   &  40.8(13) ps  \\
$^{128}$Xe  & 54 & 74 & 2+ &  442.911(9)   &  18(4) ps  \\
$^{130}$Xe  & 54 & 76 & 2+ &  536.068(6)   &  8.6(15) ps  \\
$^{132}$Xe  & 54 & 78 & 2+ &  667.715(2)   &  4.63(30) ps  \\
$^{134}$Xe  & 54 & 80 & 2+ &  847.041(23)   &  2.08(14) ps  \\
$^{136}$Xe  & 54 & 82 & 2+ &  1313.06(7)   &  0.36(14) ps  \\
$^{138}$Xe  & 54 & 84 & 2+ &  588.826(18)   &  10.5(16) ps  \\
$^{140}$Xe  & 54 & 86 & 2+ &  376.658(15)   &  70.5(20) ps  \\
$^{142}$Xe  & 54 & 88 & 2+ &  287.2(20)   &  0.20(3) ns  \\
$^{144}$Xe & 54 & 90 &  (2+)   & 252.6 &     351$^{+107}_{-67}$ ps$^*$   \\
$^{118}$Ba  & 56 & 62 &  (2+)   & 194 &       \\
$^{120}$Ba  & 56 & 64 &  (2+)   &  186(10)   &       \\
$^{122}$Ba  & 56 & 66 & 2+ &  195.9(20)   &  297(27) ps  \\
$^{124}$Ba  & 56 & 68 & 2+ &  229.91(10)   &  191(8) ps  \\
$^{126}$Ba  & 56 & 70 & 2+ &  256.02(6)   &  137(7) ps  \\
$^{128}$Ba  & 56 & 72 & 2+ &  284(8)   &  105(9) ps  \\
$^{130}$Ba  & 56 & 74 & 2+ &  357.38(8)   &  41.8(12) ps  \\
$^{132}$Ba  & 56 & 76 & 2+ &  464.508(12)   &  15.1(11) ps  \\
$^{134}$Ba  & 56 & 78 & 2+ &  604.7223(19)   &  5.12(9) ps  \\
$^{136}$Ba  & 56 & 80 & 2+ &  818.522(10)   &  1.89(3) ps  \\
$^{138}$Ba  & 56 & 82 & 2+ &  1435.805(10)   &  0.199(6) ps  \\
$^{140}$Ba  & 56 & 84 & 2+ &  602.37(3)   &  7.2(9) ps  \\
$^{142}$Ba  & 56 & 86 & 2+ &  359.596(14)   &  65(2) ps  \\
$^{144}$Ba  & 56 & 88 & 2+ &  199.326(6)   &  0.71(2) ns  \\
$^{146}$Ba  & 56 & 90 & 2+ &  181.04(5)   &  0.859(26) ns  \\
$^{148}$Ba  & 56 & 92 & 2+ &  141.8(1)   &       \\
$^{122}$Ce  & 58 & 64 &  (2+)   &  136.4(5)   &       \\
$^{124}$Ce  & 58 & 66 & 2+ &  141.9(20)   &  0.88(19) ns  \\
$^{126}$Ce  & 58 & 68 & 2+ &  169.59(3)   &  0.59(10) ns  \\
$^{128}$Ce  & 58 & 70 & 2+ &  207.09(18)   &  0.30(3) ns  \\
$^{130}$Ce  & 58 & 72 & 2+ &  253.85(16)   &  143(6) ps  \\
$^{132}$Ce  & 58 & 74 & 2+ &  325.34(8)   &  40(3) ps  \\
$^{134}$Ce  & 58 & 76 & 2+ &  409.2(10)   &  23(2) ps  \\
$^{136}$Ce  & 58 & 78 & 2+ &  552.05(13)   &  6.7(8) ps  \\
$^{138}$Ce  & 58 & 80 & 2+ &  788.744(8)   &  1.98(4) ps  \\
$^{140}$Ce  & 58 & 82 & 2+ &  1596.233(23)   &  0.091(4) ps  \\
$^{142}$Ce  & 58 & 84 & 2+ &  641.282(9)   &  5.56(12) ps  \\
$^{144}$Ce  & 58 & 86 & 2+ &  397.441(9)   &  35.4(20) ps  \\
$^{146}$Ce  & 58 & 88 & 2+ &  258.45(4)   &  0.231(26) ns  \\
$^{148}$Ce  & 58 & 90 & 2+ &  158.467(5)   &  1.01(6) ns  \\
$^{150}$Ce  & 58 & 92 & 2+ &  97(10)   &  3.3(8) ns  \\
$^{152}$Ce  & 58 & 94 & 2+ &  81.2(5)   &  2.5   ns  \\
$^{128}$Nd  & 60 & 68 & 2+ &  133.66(7)   &       \\
$^{130}$Nd  & 60 & 70 & 2+ &  159.05(14)   &  0.6(25) ns  \\
$^{132}$Nd  & 60 & 72 & 2+ &  213.16(12)   &  133(8) ps  \\
$^{134}$Nd  & 60 & 74 & 2+ &  294.17(16)   &  64.4(18) ps  \\
$^{136}$Nd  & 60 & 76 & 2+ &  373.75(16)   &   22.8$^{+32}_{-32}$ ps$^*$     \\
$^{138}$Nd  & 60 & 78 & 2+ &  520.75(17)   &       \\
$^{140}$Nd  & 60 & 80 & 2+ &  773.65(6)   &  1.4(11) ps  \\
$^{142}$Nd  & 60 & 82 & 2+ &  1575.78(10)   &  0.11(2) ps  \\
$^{144}$Nd  & 60 & 84 & 2+ &  696.561(10)   &  2.97(5) ps  \\
$^{146}$Nd  & 60 & 86 & 2+ &  453.84(3)   &  20.9(9) ps  \\
$^{148}$Nd  & 60 & 88 & 2+ &  301.705(16)   &  80(3) ps  \\
$^{150}$Nd  & 60 & 90 & 2+ &  130.21(7)   &  1.48(3) ns  \\
$^{152}$Nd  & 60 & 92 & 2+ &  72.4(5)   &  4.18(23) ns  \\
$^{154}$Nd  & 60 & 94 & 2+ &  70.8(1)   &  7.7(20) ns  \\
$^{156}$Nd  & 60 & 96 & 2+ &  67.2(2)   &       \\
$^{158}$Nd  & 60 & 98 &  (2+)   &  65.9(10)   &       \\
$^{160}$Nd  & 60 & 100 &  (2+)   &  65.2(5)   &       \\
$^{130}$Sm  & 62 & 68 &  (2+)   &  122(3)   &       \\
$^{132}$Sm  & 62 & 70 &  (2+)   &  131(1)   &       \\
$^{134}$Sm  & 62 & 72 & 2+ & 163 &  0.42(4) ns  \\
$^{136}$Sm  & 62 & 74 & 2+ &  254.92(16)   &  88(9) ps  \\
$^{138}$Sm  & 62 & 76 & 2+ &  346.71(16)   &  40(6) ps  \\
$^{140}$Sm  & 62 & 78 & 2+ &  530.68(10)   &  6.1(32) ps  \\
$^{142}$Sm  & 62 & 80 & 2+ &  768.08(19)   &    1.50$^{+22}_{-17}$ ps$^*$       \\
$^{144}$Sm  & 62 & 82 & 2+ &  1660.027(10)   &  84.4(25) fs  \\
$^{146}$Sm  & 62 & 84 & 2+ &  747.174(11)   &  $\leq$7.2   ps  \\
$^{148}$Sm  & 62 & 86 & 2+ &  550.255(8)   &  7.72(32) ps  \\
$^{150}$Sm  & 62 & 88 & 2+ &  333.955(10)   &  48.4(11) ps  \\
$^{152}$Sm  & 62 & 90 & 2+ &  121.7818(3)   &  1.403(11) ns  \\
$^{154}$Sm  & 62 & 92 & 2+ &  81.981(15)   &  3.02(4) ns  \\
$^{156}$Sm  & 62 & 94 & 2+ &  75.89(5)   &  $>$2   ns  \\
$^{158}$Sm  & 62 & 96 &  (2+)   &  72.8(10)   &       \\
$^{160}$Sm  & 62 & 98 & 2+ & 70.9 &       \\
$^{164}$Sm  & 62 & 102 &  (2+)   &  69(1)   &       \\
$^{134}$Gd  & 64 & 70 & 2+ & 115 &       \\
$^{138}$Gd  & 64 & 74 & 2+ &  220.86(18)   &  215(12) ps  \\
$^{140}$Gd  & 64 & 76 & 2+ &  328.6(3)   &       \\
$^{142}$Gd  & 64 & 78 & 2+ &  515.2(8)   &       \\
$^{144}$Gd  & 64 & 80 & 2+ &  743(17)   &       \\
$^{148}$Gd  & 64 & 84 & 2+ &  784.433(15)   &  4.2(12) ps  \\
$^{150}$Gd  & 64 & 86 & 2+ &  638.045(14)   &       \\
$^{152}$Gd  & 64 & 88 & 2+ &  344.279(12)   &  32(27) ps  \\
$^{154}$Gd  & 64 & 90 & 2+ &  123.0709(9)   &  1.184(5) ns  \\
$^{156}$Gd  & 64 & 92 & 2+ &  88.97(1)   &  2.21(2) ns  \\
$^{158}$Gd  & 64 & 94 & 2+ &  79.5143(15)   &  2.56(5) ns  \\
$^{160}$Gd  & 64 & 96 & 2+ &  75.26(1)   &  2.72(1) ns  \\
$^{162}$Gd & 64 & 98 & 2+ & 71.6 &     2758$^{+55}_{-55}$ ps$^*$      \\
$^{164}$Gd  & 64 & 100 &  (2+)   &  73.27(5)   &  2.77(14) ns  \\
$^{166}$Gd  & 64 & 102 &  (2+)   &  70(10)   &       \\
$^{140}$Dy  & 66 & 74 &  (2+)   &  202.2(20)   &       \\
$^{142}$Dy  & 66 & 76 &  (2+)   &  315.9(4)   &       \\
$^{144}$Dy  & 66 & 78 &  (2+)   &  492.5(3)   &       \\
$^{146}$Dy  & 66 & 80 & 2+ &  682.62(18)   &       \\
$^{148}$Dy  & 66 & 82 & 2+ & 1677.3 &       \\
$^{150}$Dy  & 66 & 84 & 2+ &  803.64(9)   &       \\
$^{152}$Dy  & 66 & 86 & 2+ &  613.83(5)   &  10(5) ps  \\
$^{154}$Dy  & 66 & 88 & 2+ &  334.34(3)   &  27.5(20) ps  \\
$^{156}$Dy  & 66 & 90 & 2+ &  137.77(8)   &  0.823(7) ns  \\
$^{158}$Dy  & 66 & 92 & 2+ &  98.918(10)   &  1.66(3) ns  \\
$^{160}$Dy  & 66 & 94 & 2+ &  86.7878(3)   &  2.02(1) ns  \\
$^{162}$Dy  & 66 & 96 & 2+ &  80.661(3)   &  2.19(2) ns  \\
$^{164}$Dy  & 66 & 98 & 2+ &  73.393(5)   &  2.393(29) ns  \\
$^{166}$Dy  & 66 & 100 & 2+ &  76.587(1)   &       \\
$^{170}$Dy  & 66 & 104 &  (2+)   &  71.47(15)   &       \\
$^{144}$Er  & 68 & 76 & 2+ &  330(10)   &       \\
$^{148}$Er  & 68 & 80 & 2+ &  645.89(10)   &       \\
$^{150}$Er  & 68 & 82 & 2+ &  1578.33(23)   &       \\
$^{152}$Er  & 68 & 84 & 2+ &  808.3(1)   &       \\
$^{154}$Er  & 68 & 86 & 2+ &  560.8(1)   &       \\
$^{156}$Er  & 68 & 88 & 2+ &  344.53(6)   &  34(9) ps  \\
$^{158}$Er  & 68 & 90 & 2+ &  192.15(3)   &  257(18) ps  \\
$^{160}$Er  & 68 & 92 & 2+ &  125.8(1)   &  919(31) ps  \\
$^{162}$Er  & 68 & 94 & 2+ &  102.04(3)   &  1.29(6) ns  \\
$^{164}$Er  & 68 & 96 & 2+ &  91.38(22)   &  1.569(34) ns  \\
$^{166}$Er  & 68 & 98 & 2+ &  80.5776(20)   &  1.815(23) ns  \\
$^{168}$Er  & 68 & 100 & 2+ &  79.804(1)   &  1.853(25) ns  \\
$^{170}$Er  & 68 & 102 & 2+ &  78.59(22)   &  1.896(23) ns  \\
$^{172}$Er  & 68 & 104 &  (2+)   &  77(2)   &       \\
$^{152}$Yb  & 70 & 82 & 2+ &  1531.4(5)   &       \\
$^{154}$Yb  & 70 & 84 &  (2+)   &  821.3(2)   &       \\
$^{156}$Yb  & 70 & 86 & 2+ & 536 &       \\
$^{158}$Yb  & 70 & 88 &  (2+)   &  358.2(1)   &  25(3) ps  \\
$^{160}$Yb  & 70 & 90 & 2+ &  243.1(1)   &  121(7) ps  \\
$^{162}$Yb  & 70 & 92 & 2+ &  166.72(4)   &  415(9) ps  \\
$^{164}$Yb  & 70 & 94 & 2+ &  123.31(23)   &  932(30) ps  \\
$^{166}$Yb  & 70 & 96 & 2+ &  102.37(3)   &  1.24(6) ns  \\
$^{168}$Yb  & 70 & 98 & 2+ &  87.73(1)   &  1.49(4) ns  \\
$^{170}$Yb  & 70 & 100 & 2+ &  84.25468(8)   &  1.61(2) ns  \\
$^{172}$Yb  & 70 & 102 & 2+ &  78.7427(6)   &  1.65(5) ns  \\
$^{174}$Yb  & 70 & 104 & 2+ &  76.471(1)   &  1.79(4) ns  \\
$^{176}$Yb  & 70 & 106 & 2+ &  82.135(15)   &  1.76(5) ns  \\
$^{178}$Yb  & 70 & 108 & 2+ &  84(3)   &       \\
$^{154}$Hf  & 72 & 82 &  (2+)   & 1513 &       \\
$^{156}$Hf  & 72 & 84 & 2+ & 857.2 &       \\
$^{158}$Hf  & 72 & 86 & 2+ &  476.36(11)   &       \\
$^{160}$Hf  & 72 & 88 & 2+ &  389.4(10)   &       \\
$^{162}$Hf  & 72 & 90 & 2+ & 285 &  103(8) ps  \\
$^{164}$Hf  & 72 & 92 & 2+ &  210.7(3)   &  301(29) ps  \\
$^{166}$Hf  & 72 & 94 & 2+ &  158.64(5)   &  497(23) ps  \\
$^{168}$Hf  & 72 & 96 & 2+ &  124.1(5)   &  0.89(4) ns  \\
$^{170}$Hf  & 72 & 98 & 2+ &  100.74(4)   &  1.21(4) ns  \\
$^{172}$Hf  & 72 & 100 & 2+ &  95.22(4)   &  1.55(10) ns  \\
$^{174}$Hf  & 72 & 102 & 2+ &  90.985(19)   &  1.66(7) ns  \\
$^{176}$Hf  & 72 & 104 & 2+ &  88.349(24)   &  1.43(4) ns  \\
$^{178}$Hf  & 72 & 106 & 2+ &  93.1803(10)   &  1.494(23) ns  \\
$^{180}$Hf  & 72 & 108 & 2+ &  93.324(20)   &  1.519(10) ns  \\
$^{182}$Hf  & 72 & 110 & 2+ &  97.79(9)   &       \\
$^{184}$Hf  & 72 & 112 &  (2+)   &  107.1(1)   &       \\
$^{158}$W  & 74 & 84 &  (2+)   & 913 &       \\
$^{160}$W  & 74 & 86 & 2+ &  609.9(2)   &       \\
$^{162}$W  & 74 & 88 &  (2+)   &  449.5(3)   &       \\
$^{164}$W  & 74 & 90 & 2+ &  331.9(5)   &  18(12) ps$^{\dagger}$   \\
$^{166}$W  & 74 & 92 & 2+ &  252(3)   &       \\
$^{168}$W  & 74 & 94 & 2+ &  199.3(2)   &  213(10) ps  \\
$^{170}$W  & 74 & 96 & 2+ &  156.72(13)   &  497(10) ps  \\
$^{172}$W  & 74 & 98 & 2+ &  123.2(1)   &  0.74(6) ns  \\
$^{174}$W  & 74 & 100 & 2+ &  113(1)   &  1.14(7) ns  \\
$^{176}$W & 74 & 102 & 2+ &  108.3(7)   &    992$^{+62}_{-62}$ ps$^*$      \\
$^{178}$W & 74 & 104 & 2+ &  105.9(9)   &     1138$^{+15}_{-15}$ ps$^*$        \\
$^{180}$W  & 74 & 106 & 2+ &  103.561(16)   &  1.28(5) ns  \\
$^{182}$W  & 74 & 108 & 2+ &  100.10598(7)   &  1.381(10) ns  \\
$^{184}$W  & 74 & 110 & 2+ &  111.2174(4)   &  1.251(12) ns  \\
$^{186}$W  & 74 & 112 & 2+ &  122.63(15)   &  1.036(10) ns  \\
$^{188}$W  & 74 & 114 & 2+ &  143.16(8)   &  0.87(12) ns  \\
$^{190}$W  & 74 & 116 &  (2+)   &  206.8(5)   &       \\
$^{192}$W  & 74 & 118 &  [2+]  & 219 &       \\
$^{162}$Os  & 76 & 86 &  (2+)   &  706.7(3)   &       \\
$^{164}$Os  & 76 & 88 &  (2+)   &  548(2)   &       \\
$^{166}$Os  & 76 & 90 & 2+ &  432(3)   &       \\
$^{168}$Os  & 76 & 92 & 2+ &  341.2(20)   &       \\
$^{170}$Os  & 76 & 94 & 2+ &  286.7(14)   &       \\
$^{172}$Os  & 76 & 96 & 2+ &  227.77(9)   &  116(7) ps  \\
$^{174}$Os  & 76 & 98 & 2+ &  158.6(10)   &  0.35(4) ns  \\
$^{176}$Os & 76 & 100 & 2+ &  135.1(7)   & 839$^{+125}_{-125}$ ps$^*$     \\
$^{178}$Os  & 76 & 102 & 2+ &  132.2(17)   &  0.69(5) ns  \\
$^{180}$Os  & 76 & 104 & 2+ &  132.11(10)   &  0.67(7) ns  \\
$^{182}$Os  & 76 & 106 & 2+ &  126.89(8)   &  813(11) ps  \\
$^{184}$Os  & 76 & 108 & 2+ &  119.77(9)   &  1.184(13) ns  \\
$^{186}$Os  & 76 & 110 & 2+ &  137.159(8)   &  875(15) ps  \\
$^{188}$Os  & 76 & 112 & 2+ &  155.043(4)   &  0.704(7) ns  \\
$^{190}$Os  & 76 & 114 & 2+ &  186.718(2)   &  371(8) ps  \\
$^{192}$Os  & 76 & 116 & 2+ &  205.79442(9)   &  288(4) ps  \\
$^{194}$Os  & 76 & 118 &  (2+)   &  218.509(6)   &       \\
$^{196}$Os  & 76 & 120 &  (2+)   &  324.4(10)   &       \\
$^{198}$Os  & 76 & 122 &  (2+)   &  465.4(5)   &       \\
$^{168}$Pt  & 78 & 90 &  (2+)   &  581.4(10)   &       \\
$^{170}$Pt  & 78 & 92 & 2+ &  509.2(20)   &       \\
$^{172}$Pt  & 78 & 94 &  2(+)   &  457.6(10)   &       \\
$^{174}$Pt  & 78 & 96 & 2+ &  394.2(10)   &       \\
$^{176}$Pt  & 78 & 98 & 2+ &  264(3)   &  76(7) ps  \\
$^{178}$Pt  & 78 & 100 & 2+ &  170.3(10)   & 286$^{+21}_{-21}$ ps$^*$        \\
$^{180}$Pt  & 78 & 102 & 2+ &  153.24(7)   &  374(35) ps  \\
$^{182}$Pt  & 78 & 104 & 2+ &  154.97(9)   &  479(30) ps  \\
$^{184}$Pt  & 78 & 106 & 2+ &  162.98(6)   &  360(12) ps  \\
$^{186}$Pt  & 78 & 108 & 2+ &  191.53(4)   &  260(12) ps  \\
$^{188}$Pt  & 78 & 110 & 2+ &  265.61(5)   &  66(3) ps  \\
$^{190}$Pt  & 78 & 112 & 2+ &  295.78(3)   &  62.3(31) ps  \\
$^{192}$Pt  & 78 & 114 & 2+ &  316.50645(15)   &  43.7(9) ps  \\
$^{194}$Pt  & 78 & 116 & 2+ &  328.464(5)   &  41.9(6) ps  \\
$^{196}$Pt  & 78 & 118 & 2+ &  355.6841(20)   &  34.15(15) ps  \\
$^{198}$Pt  & 78 & 120 & 2+ &  407.22(5)   &  22.25(15) ps  \\
$^{200}$Pt  & 78 & 122 & 2+ &  470.1(20)   &       \\
$^{202}$Pt  & 78 & 124 &  (2+)   &  534.9(20)   &       \\
$^{204}$Pt  & 78 & 126 &  (2+)   &  872(10)   &       \\
$^{172}$Hg  & 80 & 92 &  (2+)   &  672.8(4)   &       \\
$^{176}$Hg  & 80 & 96 & 2+ &  613.3(10)   &       \\
$^{178}$Hg  & 80 & 98 & 2+ &  558(20)   &       \\
$^{184}$Hg  & 80 & 104 & 2+ &  366.78(9)   &  21(5) ps  \\
$^{186}$Hg  & 80 & 106 & 2+ &  405.33(14)   &  18(3) ps  \\
$^{188}$Hg  & 80 & 108 & 2+ &  412.91(8)   &  13.1(20) ps  \\
$^{190}$Hg  & 80 & 110 & 2+ &  416.32(14)   &  15(1) ps$^{\dagger}$   \\
$^{192}$Hg  & 80 & 112 & 2+ &  422.79(10)   &       \\
$^{194}$Hg  & 80 & 114 & 2+ &  427.89(9)   &       \\
$^{196}$Hg  & 80 & 116 & 2+ &  425.98(10)   &  17.2(6) ps  \\
$^{198}$Hg  & 80 & 118 & 2+ &  411.80251(17)   &  23.15(28) ps  \\
$^{200}$Hg  & 80 & 120 & 2+ &  367.943(10)   &  46.4(4) ps  \\
$^{202}$Hg  & 80 & 122 & 2+ &  439.512(8)   &  27.26(22) ps  \\
$^{204}$Hg  & 80 & 124 & 2+ &  436.552(8)   &  40.3(3) ps  \\
$^{206}$Hg  & 80 & 126 & 2+ &  1068.2(20)   &  $<$21   ns  \\
$^{208}$Hg  & 80 & 128 &  (2+)   &  669(5)   &       \\
$^{210}$Hg  & 80 & 130 &  (2+)   & 643 &       \\
$^{180}$Pb  & 82 & 98 &  (2+)   &  1168(1)   &       \\
$^{182}$Pb  & 82 & 100 &  (2+)   &  888.3(3)   &       \\
$^{196}$Pb  & 82 & 114 & 2+ &  1049.2(9)   &  $<$100   ns$^{\dagger}$   \\
$^{198}$Pb  & 82 & 116 & 2+ &  1063.5(20)   &       \\
$^{200}$Pb  & 82 & 118 & 2+ &  1026.61(14)   &       \\
$^{202}$Pb  & 82 & 120 & 2+ &  960.67(5)   &  $\leq$0.1   ns  \\
$^{204}$Pb  & 82 & 122 & 2+ &  899.165(25)   &  2.88(3) ps  \\
$^{206}$Pb  & 82 & 124 & 2+ &  803.054(25)   &  8.3(25) ps  \\
$^{210}$Pb  & 82 & 128 & 2+ &  799.7(1)   &  17(5) ps  \\
$^{212}$Pb  & 82 & 130 &  (2+)   &  804.9(2)   &       \\
$^{214}$Pb  & 82 & 132 &  (2+)   &  835(1)   &       \\
$^{216}$Pb  & 82 & 134 &  (2+)   &  887(1)   &       \\
$^{190}$Po  & 84 & 106 &  (2+)   &  234.1(9)   &       \\
$^{192}$Po  & 84 & 108 &  (2+)   &  262(3)   &       \\
$^{194}$Po  & 84 & 110 & 2+ &  319.8(3)   & 25.6$^{+49}_{-49}$ ps$^*$         \\
$^{196}$Po & 84 & 112 & 2+ &  463.12(9)   & 8.0$^{+10}_{-10}$ ps$^*$      \\
$^{198}$Po & 84 & 114 & 2+ &  604.94(10)   & 2.60$^{+60}_{-49}$ ps$^*$       \\
$^{200}$Po & 84 & 116 & 2+ &  665.9(10)   & 2.00$^{+12}_{-10}$ ps$^*$       \\
$^{202}$Po  & 84 & 118 & 2+ &  677.2(20)   & 1.74$^{+52}_{-41}$ ps$^*$      \\
$^{204}$Po  & 84 & 120 & 2+ &  684.341(10)   &       \\
$^{206}$Po  & 84 & 122 & 2+ &  700.66(3)   &       \\
$^{208}$Po  & 84 & 124 & 2+ &  686.526(20)   &       \\
$^{210}$Po  & 84 & 126 & 2+ &  1181.398(10)   &  5.9(12) ps  \\
$^{212}$Po  & 84 & 128 & 2+ &  727.33(9)   &  14.2(18) ps$^{\dagger}$   \\
$^{214}$Po  & 84 & 130 & 2+ &  609.316(4)   &   $<$4 ps$^*$     \\
$^{216}$Po  & 84 & 132 & 2+ &  549.76(4)   &       \\
$^{218}$Po  & 84 & 134 & 2+ &  509.7(10)   &       \\
$^{198}$Rn  & 86 & 112 &  (2+)   &  339(2)   &       \\
$^{200}$Rn  & 86 & 114 & 2+ &  432.6(20)   &       \\
$^{202}$Rn   & 86 & 116 & 2+ &  504(10)   & 8.4$^{+30}_{-21}$ ps$^*$      \\
$^{204}$Rn   & 86 & 118 & 2+ &  542.9(10)   & 3.9$^{+17}_{-11}$ ps$^*$       \\
$^{206}$Rn  & 86 & 120 & 2+ &  575.3(10)   &       \\
$^{208}$Rn  & 86 & 122 & 2+ &  635.8(2)   &       \\
$^{210}$Rn  & 86 & 124 & 2+ &  643.9(10)   &       \\
$^{212}$Rn  & 86 & 126 & 2+ &  1273.7(10)   &       \\
$^{214}$Rn  & 86 & 128 & 2+ & 694.7 &  $<$1.4   ns  \\
$^{216}$Rn  & 86 & 130 & 2+ &  461.4(2)   &       \\
$^{218}$Rn  & 86 & 132 & 2+ &  324.32(18)   &  $<$80   ps  \\
$^{220}$Rn  & 86 & 134 & 2+ &  240.986(6)   &  0.146(5) ns  \\
$^{222}$Rn  & 86 & 136 & 2+ &  186.211(13)   &  0.32(2) ns  \\
$^{206}$Ra  & 88 & 118 &  (2+)   &  474.3(5)   &       \\
$^{208}$Ra  & 88 & 120 &  (2+)   &  520.2(2)   &       \\
$^{210}$Ra  & 88 & 122 & 2+ &  603.7(3)   &       \\
$^{212}$Ra  & 88 & 124 & 2+ &  629.3(10)   &       \\
$^{214}$Ra  & 88 & 126 & 2+ & 1382.4 &       \\
$^{216}$Ra  & 88 & 128 & 2+ &  688.2(20)   &       \\
$^{218}$Ra  & 88 & 130 & 2+ &  388.9(10)   &  29.8(28) ps  \\
$^{220}$Ra  & 88 & 132 & 2+ &  178.47(12)   &       \\
$^{222}$Ra  & 88 & 134 & 2+ &  111.12(2)   &  0.52(4) ns  \\
$^{224}$Ra  & 88 & 136 & 2+ &  84.372(3)   &  0.748(19) ns  \\
$^{226}$Ra  & 88 & 138 & 2+ &  67.67(1)   &  0.63(2) ns  \\
$^{228}$Ra  & 88 & 140 & 2+ &  63.823(20)   &  550(20) ps  \\
$^{230}$Ra  & 88 & 142 & 2+ &  57.4(1)   &       \\
$^{232}$Ra  & 88 & 144 &  (2+)   &  54.5(10)   &       \\
$^{214}$Th  & 90 & 124 &  (2+)   &  623(10)   &       \\
$^{216}$Th  & 90 & 126 & 2+ &  1478.2(1)   &       \\
$^{218}$Th  & 90 & 128 & 2+ &  689.6(6)   &       \\
$^{220}$Th  & 90 & 130 & 2+ &  386.5(10)   &       \\
$^{222}$Th  & 90 & 132 & 2+ & 183.3 &  240(20) ps  \\
$^{224}$Th  & 90 & 134 & 2+ &  98.1(3)   &  0.59(40) ns  \\
$^{226}$Th  & 90 & 136 & 2+ &  72.2(4)   &  0.395(20) ns  \\
$^{228}$Th  & 90 & 138 & 2+ &  57.773(3)   &  0.406(7) ns  \\
$^{230}$Th  & 90 & 140 & 2+ &  53.227(11)   &  0.354(9) ns  \\
$^{232}$Th  & 90 & 142 & 2+ &  49.369(9)   &  345(15) ps  \\
$^{234}$Th  & 90 & 144 & 2+ &  49.55(6)   &  0.37(3) ns  \\
$^{236}$Th  & 90 & 146 &  (2+)   &  48.4(3)   &       \\
$^{226}$U  & 92 & 134 &  (2+)   &  81.3(6)   &       \\
$^{228}$U  & 92 & 136 & 2+ &  59(14)   &       \\
$^{230}$U  & 92 & 138 & 2+ &  51.727(23)   &  0.26(3) ns  \\
$^{232}$U  & 92 & 140 & 2+ &  47.573(8)   &  245(20) ps  \\
$^{234}$U  & 92 & 142 & 2+ &  43.4981(10)   &  0.252(7) ns  \\
$^{236}$U  & 92 & 144 & 2+ &  45.244(20)   &  234(6) ps  \\
$^{238}$U  & 92 & 146 & 2+ &  44.916(13)   &  206(3) ps  \\
$^{240}$U  & 92 & 148 &  (2+)   &  45(1)   &       \\
$^{242}$U  & 92 & 150 &  (2+)   &  47.8(3)   &       \\
$^{236}$Pu  & 94 & 142 & 2+ &  44.63(10)   &       \\
$^{238}$Pu  & 94 & 144 & 2+ &  44.065(15)   &  175(3) ps  \\
$^{240}$Pu  & 94 & 146 & 2+ &  42.824(8)   &  167(6) ps  \\
$^{242}$Pu  & 94 & 148 & 2+ &  44.54(2)   &  158(3) ps  \\
$^{244}$Pu  & 94 & 150 & 2+ &  44.2(4)   &  158(11) ps  \\
$^{246}$Pu  & 94 & 152 & 2+ &  46.7(2)   &       \\
$^{236}$Cm  & 96 & 140 & 2+ & 45 &       \\
$^{238}$Cm  & 96 & 142 & 2+ &  35(7)   &       \\
$^{240}$Cm  & 96 & 144 &  (2+)   &  38(5)   &  132(9) ps  \\
$^{242}$Cm  & 96 & 146 & 2+ &  42.13(5)   &       \\
$^{244}$Cm  & 96 & 148 & 2+ &  42.957(9)   &  97(5) ps  \\
$^{246}$Cm  & 96 & 150 & 2+ &  42.852(5)   &  123(2) ps  \\
$^{248}$Cm  & 96 & 152 & 2+ &  43.4(3)   &  122.5(25) ps  \\
$^{250}$Cm  & 96 & 154 & 2+ &  43(5)   &       \\
$^{244}$Cf  & 98 & 146 & 2+ &  37(22)   &       \\
$^{246}$Cf  & 98 & 148 &  (2+)   & 44 &       \\
$^{248}$Cf  & 98 & 150 & 2+ &  41.53(6)   &       \\
$^{250}$Cf  & 98 & 152 & 2+ &  42.721(5)   &  96(10) ps  \\
$^{252}$Cf  & 98 & 154 & 2+ &  45.72(5)   &  92(6) ps  \\
$^{254}$Cf  & 98 & 156 &  (2+)  & 50 &       \\
$^{248}$Fm  & 100 & 148 & 2+ &  46(1)   &       \\
$^{252}$Fm  & 100 & 152 & 2+ &  42.1(13)   &       \\
$^{254}$Fm  & 100 & 154 & 2+ &  44.992(10)   &       \\
$^{256}$Fm  & 100 & 156 & 2+ &  48.12(16)   &       \\
$^{252}$No  & 102 & 150 &  (2+)   &  46.4(10)   &       \\
$^{254}$No  & 102 & 152 & 2+ &  44.2(4)   &       \\
$^{256}$Rf  & 104 & 152 &  (2+)   &  44(1)   &       \\
 \label{table:2+}      
\end{longtable}
\end{center}

\section{Conclusions}

Spin and parity assignments in even-even nuclei have been a fascinated topic in nuclear physics for the last 70 years. Many new measurements have been conducted in recent years~\cite{NRC2013}, and the data update was a long time overdue. We surveyed first-excited state properties across the nuclear chart using the Evaluated Nuclear Structure Data File (ENSDF) and other available data.  The prevalence of 2$^+_1$  states was confirmed, and properties of   0$^+_2$, 1$^-_1$, and 3$^-_1$  states were reevaluated.   

In summary, we would reiterate that there is no comprehensive theoretical explanation for the 2$^+$ lowest excited state spin and parity dominance in even-even nuclei, and previous theoretical works~\cite{Bel62,Gold76}  imply that both neutron and protons should be considered as contributing factors.  We hope the current nuclear properties update of   2$^+_1$  in conjunction with 0$^+_2$, 1$^-_1$, and 3$^-_1$  states would stimulate future theoretical and experimental studies that would help to clarify this phenomenon.

\section{Acknowledgments}
\label{sec:Acknowledgements}
The authors are indebted to the International Network of Nuclear Structure and Decay Data  (NSDD) Evaluators~\cite{NSDD,Dim20}  members for their tireless work on the ENSDF library evaluations,  Vladimir Zelevinsky (Michigan State University) for useful discussions, and the referee for good comments and productive suggestions that helped to improve the manuscript. 
Work at Brookhaven was funded by the Office of Nuclear Physics, Office of Science of the U.S. Department
of Energy, under Contract No. DE-SC0012704 with Brookhaven Science Associates, LLC.  \\ \\




\end{document}